\documentclass[12pt]{article}
\usepackage{chetnew}
\usepackage{xparse}
\usepackage{amssymb, amsmath}
\usepackage{xcolor}
\usepackage{tikz}
\usepackage{cite}
\usepackage[linktocpage]{hyperref}
\usepackage{mathtools}
\usepackage{physics}

\DeclareFontShape{OT1}{cmr}{mx}{n}%
    {<->cmr10}{}

\renewcommand{\ket}[1]{|#1\rangle}

\renewcommand{\tilde}{\widetilde}

\allowdisplaybreaks

\preprint{
NITEP 199}

\title{Liouville Irregular States of Half-Integer Ranks}

\author{Ryo Hamachika$^{1\spadesuit}$, Tomoki Nakanishi$^{1\diamondsuit}$, Takahiro
Nishinaka$^{1,2,3\clubsuit}$ and Shou Tanigawa$^{1\heartsuit}$}
\affiliation{\smallskip  $^1$Department of Physics, Graduate School of Science\\
Osaka Metropolitan University, Osaka 558-8585, Japan\\
\bigskip
$^2$Nambu Yoichiro Institute of Theoretical and Experimental Physics (NITEP)\\
Osaka Metropolitan University, Osaka 558-8585, Japan\\
\bigskip
$^3$Osaka Central Advanced Mathematical Institute (OCAMI)\\
Osaka Metropolitan University, Osaka 558-8585, Japan
\emails{$^{\spadesuit}$sf22775d@st.omu.ac.jp, $^{\diamondsuit}$nakanishiphys@gmail.com,
$^{\clubsuit}$nishinaka@omu.ac.jp, $^{\heartsuit}$sf22851y@st.omu.ac.jp}
}

\abstract{
We conjecture a set of differential equations that characterizes
the Liouville irregular states of half-integer ranks, which extends the
generalized AGT correspondence to all the
 $(A_1,A_\text{even})$ and $(A_1,D_\text{odd})$ types Argyres-Douglas
theories.  For lower
half-integer ranks, our conjecture is verified by deriving it as a 
suitable
limit of a similar set of differential equations for integer
ranks. This limit is interpreted as the 2D
counterpart of a 4D RG-flow from $(A_1,D_{2n})$ to $(A_1,D_{2n-1})$. For
rank $3/2$, we solve the conjectured differential equations and find a
power series expression for the irregular state $|I^{(3/2)}\rangle$. 
For rank $5/2$, our conjecture is consistent with the differential equations
recently discovered by H. Poghosyan and R. Poghossian.
}

\begin{document}

\maketitle
\toc

\section{Introduction}
\label{sec:intro}

The AGT correspondence \cite{Alday:2009aq, Wyllard:2009hg,
Gaiotto:2009ma} and its
generalizations revealed a surprising connection between
four-dimensional $\mathcal{N}=2$ supersymmetric quantum field theories
(QFTs) and two-dimensional conformal field theories (CFTs). Among other
things, one important application of this correspondence is to evaluate the partition
functions of various strongly coupled 4D
$\mathcal{N}=2$ QFTs without Lagrangian description. While these
partition functions cannot be directly evaluated by supersymmetric
localization, the AGT correspondence implies that one can evaluate them as correlation functions
of 2D CFTs.

One of the most well-studied classes of non-Lagrangian 4D
$\mathcal{N}=2$ QFTs in the above context is Argyres-Douglas (AD) superconformal field theories
(SCFTs). These SCFTs are realized at a special
singular point on the Coulomb branch of $\mathcal{N}=2$ gauge theories
\cite{Argyres:1995jj, Argyres:1995xn, Eguchi:1996vu, Argyres:2012fu} and also realized by type IIB string theory on singular Calabi-Yau
three-folds \cite{Shapere:1999xr, Cecotti:2010fi}, but in this paper we focus on their realization
on wrapped M5-branes \cite{Bonelli:2011aa, Xie:2012hs}. To be more specific, we compactify two M5-branes
on a sphere with an irregular puncture with (or without) a
regular puncture, which gives rise to an AD theory called
$(A_1,D_{2r})$ (or $(A_1,A_{2r-3})$) theory. Here, $r$ is an integer or
a half-integer that characterizes the irregular puncture, and is called the
``rank'' of the puncture. The generalized AGT correspondence
\cite{Bonelli:2011aa, Gaiotto:2012sf}
then implies that the Nekrasov partition function of these
$(A_1,D_{2r})$ and $(A_1,A_{2r-3})$ theories are evaluated as  two- and one-point functions of
the 2D Liouville CFT, respectively. By the state-operator map, these correlation
functions are equivalent to the inner products of states, and therefore the
partition functions of these AD theories are identified as
\begin{align}
 \mathcal{Z}_{(A_1,D_{2r})} = \langle \Delta |I^{(r)}\rangle~, \qquad
 \mathcal{Z}_{(A_1,A_{2r-3})} =\langle 0|I^{(r)}\rangle~.
\label{eq:AGT}
\end{align} 
Here, $|0\rangle$ is the vacuum, $|\Delta\rangle$ is a primary state
so that $L_0|\Delta \rangle = \Delta |\Delta\rangle$, and $|I^{(r)}\rangle$ is a non-primary state corresponding
the irregular puncture of rank $r$. See \cite{Kanno:2012xt, Nishinaka:2012kn, Rim:2012tf, Kanno:2013vi, Matsuo:2014rba, 
Nagoya:2015cja, Choi:2015idw,  Rim:2016msx,Nagoya:2016mlj, Nagoya:2018,
Itoyama:2018wbh, Itoyama:2018gnh,Nishinaka:2019nuy, Kimura:2020krd,
Kimura:2022yua, Fucito:2023plp, Poghosyan:2023zvy} for recent works on
the irregular conformal blocks related to
Argyres-Douglas theories.

According to \cite{Bonelli:2011aa, Xie:2012hs}, the rank $r$ of the irregular puncture can
generally be an integer or a half-integer. When $r$ is an integer, the
complete characterization of the irregular state $|I^{(r)}\rangle$ was
found in \cite{Gaiotto:2012sf} after the pioneering works on lower ranks
\cite{Gaiotto:2009ma, Bonelli:2011aa}. Namely, for a positive integer $n$,
the irregular state $|I^{(n)}\rangle$ is a simultaneous
solution to the following set of differential equations:
\begin{align}
 L_k|I^{(n)}\rangle =\left\{
\begin{array}{l} 
0 \quad \text{for}\quad 2n<k
\\[2mm]
\Lambda_k |I^{(n)}\rangle \quad \text{for}\quad n\leq k \leq 2n
\\[2mm]
\left(\Lambda_k + \sum_{\ell=1}^{n-k} \ell\, c_{\ell +
 k}\frac{\partial}{\partial c_\ell}\right)|I^{(n)}\rangle \quad
\text{for}\quad 0\leq k \leq n-1
\end{array}
\right.~,
\label{eq:diff-eq-int}
\end{align}
where $\Lambda_0,\cdots, \Lambda_{2n}$ are determined by $c_0,\cdots,
c_n$ as
\begin{align}
 \Lambda_k \equiv 
\left\{
\begin{array}{l}
-\sum_{i=k-n}^{n}c_ic_{k-i} \quad \text{for}\quad n<k\leq 2n
\\[2mm]
-\sum_{i=0}^kc_{i}c_{k-i} + (k+1)Qc_k\quad \text{for}\quad k\leq n
\end{array}
\right.~.
\end{align}
It was then demonstrated in \cite{Gaiotto:2012sf, Nishinaka:2019nuy} that, when $n$ is an integer, a power series
expression for $|I^{(n)}\rangle$ can be obtained by solving \eqref{eq:diff-eq-int}
order by order of $c_n$. 

However, when the rank $r$ is a general half-integer, similar differential equations for
$|I^{(r)}\rangle$ have not been known, and therefore 
the 
Liouville
irregular state $|I^{(r)}\rangle$ for a general half-integer $r$ is still to be understood.
This means that the AGT
correspondence \eqref{eq:AGT} for $(A_1,D_{\text{odd}})$ and $(A_1,A_{\text{even}})$
has not completely been established.

Recently, one important progress on this problem has been made in \cite{Poghosyan:2023zvy}, where the
authors discovered a remarkable set of differential equations that the
rank-$5/2$ irregular state $|I^{(5/2)}\rangle$
satisfies. Moreover, it was demonstrated in \cite{Poghosyan:2023zvy} that a power series
expression for $|I^{(5/2)}\rangle$ can be obtained by solving these
equations. It would then be highly desirable to generalize their
results
to $|I^{(r)}\rangle$ for all half-integers $r$.

In this paper, we conjecture a set of differential equations that
$|I^{(n-\frac{1}{2})}\rangle$ satisfies for arbitrary positive integer $n$. To that
end, we first focus on lower values of $n$ and derive differential equations for
$|I^{(n-\frac{1}{2})}\rangle$ from those for $|I^{(n)}\rangle$. This is
done by finding a special limit from $|I^{(n)}\rangle$ to
$|I^{(n-\frac{1}{2})}\rangle$, which corresponds to a renormalization
group (RG) flow from $(A_1,D_{2n})$ to $(A_1,D_{2n-1})$. We
explicitly find such
a limit for $n=1,2,3,\cdots,7$.  This then gives us differential
equations for $|I^{(n-\frac{1}{2})}\rangle$ for these values of $n$.
The resulting equations for these seven cases turn out to be sufficient for us to
conjecture a general formula for differential equations satisfied by
$|I^{(n-\frac{1}{2})}\rangle$.
 Indeed, we conjecture that, for any positive integer $n$, the irregular state
$|I^{(n-\frac{1}{2})}\rangle$ is a simultaneous solution to 
\begin{align}
 L_k|I^{(n-\frac{1}{2})}\rangle = \left\{
\begin{array}{l}
0 \qquad (2n\leq k)
\\[3mm]
\Lambda_k|I^{(n-\frac{1}{2})}\rangle\qquad (n\leq k\leq 2n-1)
\\[3mm]
{\displaystyle \bigg(f_k(\Lambda_n,\Lambda_{n+1},\cdots,\Lambda_{2n-1}) +
 \sum_{m=n}^{2n-k-1}(m-k)\Lambda_{m+k}\frac{\partial}{\partial \Lambda_m}\bigg)|I^{(n-\frac{1}{2})}\rangle\qquad (0\leq k\leq n-1)}
\end{array}
\right.~.
\label{eq:L-halfint}
\end{align}
Here, the complex numbers $\Lambda_n,\cdots, \Lambda_{2n-1}$ are independent eigenvalues of
$L_n,\cdots,L_{2n-1}$, respectively, and $f_0(\Lambda),\cdots,
f_{n-1}(\Lambda)$ are Laurent polynomials of these eigenvalues. The
precise definition of $f_0(\Lambda),\cdots,f_{n-1}(\Lambda)$ is given in
Eq.~\eqref{ourrep} in Sec.~\ref{sec:formula}, and the explicit expressions
for $f_k(\Lambda)$ for lower values of $n$ are shown in Appendix
\ref{app:f}. Our conjectured formula \eqref{eq:L-halfint} extends the
(generalized) AGT correspondence to all the $(A_1,D_{\text{odd}})$ and
$(A_1,A_{\text{even}})$ theories, and provides a strong tool to compute
the Nekrasov partition function of these AD theories.

When $n=3$ (corresponding to rank $5/2$), our formula
\eqref{eq:L-halfint} is shown to reproduce the differential
equations obtained in  \cite{Poghosyan:2023zvy} (after renormalizing
the state). When
$n=2$ (corresponding to rank $3/2$), our conjecture gives a power-series expression for
\begin{align}
 \mathcal{Z}_{(A_1,D_3)} = \langle \Delta |I^{(3/2)}\rangle~.
\label{eq:A1D3}
\end{align}
Since the $(A_1,D_3)$ theory is equivalent to the $(A_1,A_3)$
theory, this must be identical to
\begin{align}
 \mathcal{Z}_{(A_1,A_3)} = \langle 0 |I^{(3)}\rangle~.
\label{eq:A1A3}
\end{align}
We check that the power-series expression for \eqref{eq:A1D3}
obtained by using our conjectured formula is in perfect agreement with
an expression for \eqref{eq:A1A3}
evaluated in \cite{Nishinaka:2019nuy}. 

The rest of this paper is organized as follows. In
sec.~\ref{sec:review}, we give a brief review of the Liouville irregular
state of integer ranks and their role in the AGT correspondence.
In Sec.~\ref{sec:lower-ranks}, we derive the differential equations for
$|I^{(n-\frac{1}{2})}\rangle$ from that for $|I^{(n)}\rangle$, for $n=2$
and $n=3$. Our strategy there is to consider the limit of parameters
that corresponds to an RG-flow between two AD theories in four
dimensions. In Sec.~\ref{sec:formula}, we give our conjecture on the
differential equations for $|I^{(r)}\rangle$ for a general half-integer
rank $r$. In Sec.~\ref{sec:example}, we solve our conjectured
differential equations for $|I^{(3/2)}\rangle$, and find a power series
expression for $\mathcal{Z}_{(A_1,D_3)} = \langle \Delta
|I^{(3/2)}\rangle$. We show that this power series is identical to the
power series expansion of $\mathcal{Z}_{(A_1,A_3)}=\langle
0|I^{(3)}\rangle$ evaluated in \cite{Nishinaka:2019nuy}. In
Sec.~\ref{sec:conclusion}, we conclude and discuss possible future
directions. In Appendix \ref{app:f}, we list explicit expressions for
$f_k(\Lambda_{n},\cdots,\Lambda_{2n-1})$ for lower values of $n$.



\section{Irregular states of integer ranks}
\label{sec:review}

\subsection{Recursive construction of irregular states}

In this section, we give a brief review of the irregular states of integer ranks
and their connection to the Nekrasov partition function of
Argyres-Douglas theories, following \cite{Gaiotto:2012sf, Nishinaka:2019nuy}.

As mentioned in Sec.~1, an irregular state of rank $n\in\mathbb{N}$ is
a state $|I^{(n)}\rangle$ satisfying the differential equations
\eqref{eq:diff-eq-int}, which we reproduce here for convenience:
\begin{align}
 L_k|I^{(n)}\rangle =\left\{
\begin{array}{l} 
0 \quad \text{for}\quad 2n<k
\\[2mm]
\Lambda_k |I^{(n)}\rangle \quad \text{for}\quad n\leq k \leq 2n
\\[2mm]
\left(\Lambda_k + \sum_{\ell=1}^{n-k} \ell\, c_{\ell +
 k}\frac{\partial}{\partial c_\ell}\right)|I^{(n)}\rangle \quad
\text{for}\quad 0\leq k \leq n-1
\end{array}
\right.~,
\label{eq:diff-eq-int2}
\end{align}
where $\Lambda_0,\cdots, \Lambda_{2n}$ are determined by $c_0,\cdots,
c_n$ as
\begin{align}
 \Lambda_k \equiv 
\left\{
\begin{array}{l}
-\sum_{i=k-n}^{n}c_ic_{k-i} \quad \text{for}\quad n<k\leq 2n
\\[2mm]
-\sum_{i=0}^kc_{i}c_{k-i} + (k+1)Qc_k\quad \text{for}\quad k\leq n
\end{array}
\right.~.
\end{align}
Note that these equations contain $(n+1)$ parameters,
$c_0$ and $\pmb{c} = (c_1,\,\cdots,\, c_n)$. In addition to them,
a solution to these equations depends on $n$ extra parameters
corresponding to the ``initial conditions'' for the $n$ differential
equations involved in \eqref{eq:diff-eq-int2}. We denote these extra parameters by $\pmb{\beta} =
(\beta_0,\cdots,\beta_{n-1})$ as in \cite{Nishinaka:2019nuy}

It was conjectured in
\cite{Gaiotto:2012sf} that the irregular state $|I^{(n)}\rangle$ is
uniquely fixed by specifying these $(2n+1)$ parameters, $c_0$, $\pmb{c}$ and
$\pmb{\beta}$, up to a constant prefactor. 
We therefore write
$|I^{(n)}\rangle$ as
$|I^{(n)}(c_0,\pmb{c};\pmb{\beta})\rangle$.
To be more precise, it was
conjectured in \cite{Gaiotto:2012sf} that the rank-$n$ irregular state
$|I^{(n)}\rangle$ is expanded uniquely as
\begin{align}
 |I^{(n)}(c_0,\pmb{c};\pmb{\beta})\rangle = \mathcal{M}(c_0,\pmb{c},\pmb{\beta})\sum_{k=0}^\infty
 (c_n)^k
 |I^{(n-1)}_k(c_0,\widetilde{\pmb{c}};\pmb{\beta})\rangle~,
\label{eq:exp-integer}
\end{align}
where $\widetilde{\pmb{c}} \equiv (c_1,\cdots,c_{n-1})$.
The states $|I^{(n-1)}_k(c_0, \widetilde{\pmb{c}},\pmb{\beta})\rangle$ are linear
combinations of states of the form
\begin{align}
 (L_{-y_1})^{m_1}\cdots (L_{-y_k})^{m_k}\left(\frac{\partial}{\partial
 c_1}\right)^{\ell_1}\cdots\left(\frac{\partial}{\partial c_{n-1}}\right)^{\ell_{n-1}}|I^{(n-1)}(\beta_{n-1},\widetilde{\pmb{c}};\widetilde{\pmb{\beta}})\rangle~,
\end{align}
where
$\widetilde{\pmb{\beta}}\equiv(\beta_0,\cdots\beta_{n-2})$,
$y_1>\cdots>y_k>0$, $m_i$ are positive integers, and $\ell_j$ are non-negative integers. Such
linear combinations are called ``generalized descendants'' of the
rank-$(n-1)$ state
$|I^{(n-1)}(\beta_{n-1},\widetilde{\pmb{c}};\widetilde{\pmb{\beta}})\rangle$. 

Note
that, since  $|I^{(n-1)}(\beta_{n-1},\widetilde{\pmb{c}};\widetilde{\pmb{\beta}})\rangle$ is independent of $c_n$ and
$c_0$,
identifying all $|I^{(n-1)}_k\rangle$ in the expansion \eqref{eq:exp-integer} completely
characterizes the dependence of $|I^{(n)}(c_0,\pmb{c};\pmb{\beta})\rangle$
on $c_{n}$ and $c_0$. Similarly, 
$|I^{(n-1)}(\beta_{n-1},\widetilde{\pmb{c}};\widetilde{\pmb{\beta}})\rangle$
is expected to be expanded in terms of generalized descendants of a
rank-$(n-2)$ state. This expansion completely characterizes the dependence
of
$|I^{(n-1)}(\beta_{n-1},\widetilde{\pmb{c}};\widetilde{\pmb{\beta}})\rangle$
on $c_{n-1}$ and $\beta_{n-1}$. By repeating this procedure, the rank-$n$
state is ultimately constructed from a rank-zero state
$|I^{(0)}(\beta_0)\rangle$, which is identified as a Virasoro primary
state $|\Delta_{\beta_0}\rangle$ with the highest weight
$\Delta_{\beta_0}\equiv \beta_0(Q-\beta_0)$. This implies
that $\beta_0$ specifies the highest weight of the Verma module to which
$|I^{(n)}(c_0,\pmb{c};\pmb{\beta})\rangle$ belongs.

\subsection{Parameter identification between 4D and 2D}

Let us now review how the $(2n+1)$ parameters of the irregular state
$|I^{(n)}(c_0,\pmb{c};\pmb{\beta})\rangle$ are related to the parameters
of the corresponding AD theory in four dimensions.

To that end, we focus on the $(A_1,D_{2n})$ theory whose Seiberg-Witten
(SW) curve is given by
\begin{align}
 x^2 = z^{2n-2} + \sum_{i=1}^{n-1}d_iz^{2n-2-i}  + Mz^{n-2} + \sum_{i=1}^{n-1}u_i
 z^{n-2-i} +  \frac{m^2}{z^2}~,
\label{eq:curve_A1D2n}
\end{align}
where $d_1,\cdots,d_{n-2}$ and $d_{n-1}$ are relevant couplings, $M$ and
$m$ are mass parameters, and $u_1,\cdots,u_{n-2}$ and $u_{n-1}$ are the
vacuum expectation values (VEVs) of Coulomb branch operators. The
scaling dimensions of these parameters are evaluated as
$[d_i] = i/n,\, [u_i] = 1+i/n$ and $[M]=[m] =1$. This
follows from the fact that the SW 1-form $\lambda\equiv xdz$ has scaling
dimension one.

The AGT correspondence implies that the partition function of this
theory is related to the inner product of a rank-$n$ irregular state and
a primary state as
\begin{align}
 \mathcal{Z}_{(A_1,D_{2n})} = \langle \Delta | I^{(n)}(c_0,\pmb{c};\pmb{\beta})\rangle~.
\end{align}
Note that, for the right-hand side to be non-vanishing, $|I^{(n)}\rangle$
must be in the same Verma module as $|\Delta\rangle$. This fixes the
value of $\Delta$ as 
\begin{align}
 \Delta = \Delta_{\beta_0}~,
\label{eq:Delta}
\end{align}
in terms of $\beta_0$.
The SW curve of
the four-dimensional theory is also written in terms of the irregular
state as \cite{Alday:2009aq}
\begin{align}
 x^2 = - \frac{\langle \Delta | T(z)|I^{(n)}\rangle }{\langle \Delta I^{(n)}\rangle}~,
\end{align}
where $T(z) = \sum_{k\in \mathbb{Z}}L_k/z^{k+2}$ is the 2D stress tensor.
 From \eqref{eq:diff-eq-int2} and $\langle \Delta | L_k =0$ for
$k<0$, we see that this curve is expressed as \cite{Bonelli:2011aa, Gaiotto:2012sf}
\begin{align}
 x^2 = -\frac{\Lambda_{2n}}{z^{2n+2}} - \frac{\Lambda_{2n-1}}{z^{2n+1}}-
 \cdots - \frac{\Lambda_n}{z^{n+2}}-
 \sum_{i=1}^{n-1}\frac{v_{i}}{z^{n+2-i}} - \frac{\Delta}{z^2}~,
\label{eq:AGT-curve-integer}
\end{align}
where 
\begin{align}
v_i \equiv \frac{\langle \Delta |L_{n-i} |I^{(n)}\rangle}{\langle
 \Delta|I^{(n)}\rangle}~,
\label{eq:vi}
\end{align}
for $i=1,\cdots,n-1$.  
By changing variables as $z\to 1/z$ and $x\to -z^2x$, we can rewrite
this curve into
\begin{align}
 x^2 = -\Lambda_{2n} z^{2n-2} - \Lambda_{2n-1}z^{2n-3}-\cdots- \Lambda_n
 z^{n-2} - \sum_{i=1}^{n-1}v_iz^{n-2-i} -\frac{\Delta}{z^2}
\label{eq:curve-1}
\end{align}
without changing the expression for the SW 1-form $\lambda =xdz$. By
rescaling $z$ and $x$ further as $z \to (-\Lambda_{2n})^{-\frac{1}{2n}}z$ and $x \to (-\Lambda_{2n})^{\frac{1}{2n}}x$,
this curve is further rewritten as
\begin{align}
 x^2 = z^{2n-2} - \sum_{i=1}^{n-1}\frac{\Lambda_{2n-i}}{(-\Lambda_{2n})^{\frac{2n-i}{2n}}}z^{2n-2-i} -
 \frac{\Lambda_n}{(-\Lambda_{2n})^{\frac{n}{2n}}} z^{n-2} - \sum_{i=1}^{n-1}\frac{v_i}{(-\Lambda_{2n})^{\frac{n-i}{2n}}}z^{n-2-i} -
 \frac{\Delta}{z^2}~,
\label{eq:curve-2}
\end{align}
which is indeed of the same form as the
curve 
\eqref{eq:curve_A1D2n} of the $(A_1,D_{2n})$ theory. Note that the above
changes of variables keep the SW 1-form $\lambda = xdz$ invariant.

From the identification of \eqref{eq:curve_A1D2n} and \eqref{eq:curve-2}, one can make a
precise identification of the 4D and 2D parameters:
\begin{align}
 d_i = -\frac{\Lambda_{2n-i}}{(-\Lambda_{2n})^{\frac{2n-i}{2n}}}~,\qquad
 u_i = -\frac{v_i}{(-\Lambda_{2n})^{\frac{n-i}{2n}}}~,\qquad M =
 -\frac{\Lambda_n}{\sqrt{-\Lambda_{2n}}}~,\qquad m=\sqrt{-\Delta}~.
\label{eq:parameters-4D2D}
\end{align}
Note that, since $\Lambda_k$ are completely fixed by $c_0,\cdots, c_n$, the
relevant couplings $d_i$ (and a mass parameter) in four dimensions are related to the parameters that appear in
the differential equations \eqref{eq:diff-eq-int2}. In contrast, the
VEVs of the Coulomb branch operators $u_i$ in four dimensions are
translated into $v_i$ and therefore related to the parameters
$\beta_1, \cdots, \beta_{n-1}$. These latter parameters do not appear in the
differential equations
\eqref{eq:diff-eq-int2}, but are rather ``initial
conditions'' that we need to fix when solving the equations
\eqref{eq:diff-eq-int2}.

The above interpretation is naturally understood from the viewpoint of
the class S construction of the $(A_1,D_{2n})$ theory. Indeed, the
$(A_1,D_{2n})$ theory is realized by compactifying two M5-branes on
sphere with two punctures. One of these punctures is a regular
puncture, around which the behavior of the M5-branes is characterized by the parameter $\Delta$. The other puncture is
an irregular puncture, around which the M5-branes is characterized by
$c_0,\cdots,c_n$. Therefore, the 4D relevant couplings and mass
parameters are specified by the boundary conditions of the M5-branes
near the punctures. 
This is natural since couplings and masses need to be
fixed when specifying the 4D theory by fixing the boundary behaviors of
the M5-branes. 
In contrast, 
the VEVs of the Coulomb
branch operators are {\it not} fixed just by specifying the 4D theory;
they are rather parameters specifying the vacuum of the 4D theory on the
Coulomb branch. Therefore, these VEVs are not fixed by the boundary
conditions of M5-branes, but are related to the moduli parameters of
BPS M5-branes that are now encoded in $\beta_1, \cdots, \beta_{n-1}$.

Let us also mention that, while the SW curve has only $2n$
parameters, there are $(2n+1)$ parameters on the 2D side; $c_0,\cdots,c_n$ and $\pmb{\beta} =
(\beta_0,\cdots,\beta_{n-1})$.\footnote{Recall here that $\Delta$ is
already fixed by $\beta_0$ as in \eqref{eq:Delta}.} This implies that one combination of the
$(2n+1)$ parameters on the 2D side is irrelevant on the 4D side. This
reflects the fact that the $(A_1,D_{2n})$ theory has conformal
invariance and therefore the dilatation preserves the SW curve
\eqref{eq:curve_A1D2n}. Indeed, for $\zeta \in \mathbb{R}_{+}$ the following transformation preserves
the curve:
\begin{align}
 d_i \to \zeta^{[d_i]} d_i~,\qquad u_i \to \zeta^{[u_i]}u_i~,\qquad M\to
 \zeta M~,\qquad m\to \zeta m~,
\label{eq:parameter-id}
\end{align}
up to the change of coordinates $z\to \zeta^{\frac{1}{n}}z$ and $x \to
\zeta^{1-\frac{1}{n}} x$. This conformal invariance then implies that
the coefficient of $z^{2n-2}$ in \eqref{eq:curve-1} can be set to one as
in \eqref{eq:curve-2} by
a change of coordinates. As a result, the SW
curve in four dimensions depends only on $2n$ combinations of the
$(2n+1)$ parameters in two dimensions.

\section{Irregular states of lower half-integer ranks}
\label{sec:lower-ranks}

In this section, we consider a limit from the irregular state
$|I^{(n)}\rangle$ of integer
rank $n$ to the irregular state $|I^{(n-\frac{1}{2})}\rangle$ of
half-integer rank $(n-\frac{1}{2})$, and identify differential equations
that $|I^{(n-\frac{1}{2})}\rangle$ satisfies. 
Since 
\begin{align}
 \mathcal{Z}_{(A_1,D_{2n})} =
\langle \Delta |I^{(n)}\rangle~,\qquad \mathcal{Z}_{(A_1,D_{2n-1})} = \langle \Delta |I^{(n-\frac{1}{2})}\rangle~,
\end{align}
such a limit corresponds to a limit from $(A_1,D_{2n})$ to
$(A_1,D_{2n-1})$ in four dimensions. A natural candidate for such a
limit is the RG-flow. 
Therefore, we first study the RG-flow between these two AD theories, and
then interpret it on the Liouville side.\footnote{One can also discuss a similar RG-flow from $(A_1,A_{2n-3})$
to $(A_1,A_{2n-4})$.}

\subsection{4D RG-flow}

Such RG-flows are well-studied in the literature
\cite{Shapere:1999xr, Xie:2013jc, Buican:2014qla}. Indeed, turning on the relevant coupling of the lowest
dimension in $(A_1,D_{2n})$ triggers an RG-flow whose IR
end-point is $(A_1,D_{2n-1})$. 

The easiest way to see this is to look at
the
SW curve of the $(A_1,D_{2n})$ theory shown in \eqref{eq:curve_A1D2n}.
In this curve, we make the relevant coupling $d_1$
 very large compared to the other parameters. This corresponds to the
 RG-flow triggered by $d_1\neq 0$.
 To describe the curve of the IR end-point of the RG-flow, we rescale the
coordinate as $x\to (d_1)^{\frac{1}{2n-1}}x$ and $z\to
 (d_1)^{-\frac{1}{2n-1}}z$. Note that this
rescaling preserves the SW 1-form. We also define rescaled deformation
 parameters as $\tilde{d}_{i-1} \equiv (d_1)^{-\frac{2n-i}{2n-1}}d_i$ for
 $i\geq 2$,\, $\tilde{d}_{n-1} \equiv (d_1)^{-\frac{n}{2n-1}}M$,\,
and $\tilde{u}_i \equiv  (d_1)^{-\frac{n-i}{2n-1}}u_i$ for $i\geq 1$.
Then the curve is expressed as
\begin{align}
 x^2 = \frac{1}{(d_1)^{\frac{2n}{2n-1}}}z^{2n-2} + z^{2n-3}  +
 \sum_{i=1}^{n-1}\tilde{d}_iz^{2n-3-i} + \sum_{i=1}^{n-1}\tilde{u}_i z^{n-2-i} + 
 \frac{m^2}{z^2}~.
\end{align}
We see that the limit $d_1\to \infty$ with $\tilde{d}_i,\,\tilde{u}_i$
and $m$ kept fixed gives us
\begin{align}
 x^2 = z^{2n-3}  +
 \sum_{i=1}^{n-1}\tilde{d}_iz^{2n-3-i} + \sum_{i=1}^{n-1}\tilde{u}_i z^{n-2-i} + 
 \frac{m^2}{z^2}~,
\label{eq:IR-curve}
\end{align}
which is precisely identical to the curve of the $(A_1,D_{2n-1})$ theory. 

Note
that, from $[\tilde{d}_i] = i/(n-1/2),\,
[\tilde{u}_i]=1+(i-1/2)/(n-1/2)$ and $[m]=1$, we see that the IR CFT has only one
mass parameter $m$, while the curve \eqref{eq:curve_A1D2n} of the UV theory has two
mass parameters. This means that one of the
two mass parameters of the UV CFT is lost along this
RG-flow.\footnote{Note that the lost mass parameter $M$
corresponds to the flavor moment map of an abelian flavor
symmetry. While non-abelian flavor symmetry is preserved along the
RG-flow triggered by $\mathcal{N}=2$ preserving relevant deformations,
an abelian flavor symmetry can be lost \cite{Argyres:1995xn}.}
In contrast, the number of the independent VEVs
of Coulomb branch operators has not been changed by this RG-flow. This
suggests that the differential equations for
$|I^{(n-\frac{1}{2})}\rangle$ involve the same number of independent
differential operators as in the equations \eqref{eq:diff-eq-int2} for
$|I^{(n)}\rangle$. Below, we will find such
differential equations for $|I^{(n-\frac{1}{2})}\rangle$ by interpreting
the above 4D RG-flow on the 2D side.

\subsection{2D side of the RG-flow}

We now interpret the above 4D RG-flow between $(A_1,D_{2n})$ and
$(A_1,D_{2n-1})$ on the 2D side. To that end, recall that
the SW curve \eqref{eq:curve_A1D2n} of $(A_1,D_{2n})$ theory is
identified as 
$x^2 = -\langle \Delta|T(z) |I^{(n)}\rangle/\langle \Delta
|I^{(n)}\rangle$, where $T(z) = \sum_{k\in \mathbb{Z}} L_k/z^{k+2}$ is the stress
tensor. This enables us to express the SW-curve in terms of the 2D
parameters as shown in \eqref{eq:curve-1}. Let us now take the limit $\Lambda_{2n} \to 0$
with the other parameters kept fixed. Then the curve reduces to
\begin{align}
 x^2 = -\Lambda_{2n-1}z^{2n-3}-\cdots- \Lambda_n
 z^{n-2} - \sum_{i=1}^{n-1}v_iz^{n-2-i} -\frac{\Delta}{z^2}~.
\label{eq:curve-4}
\end{align}
We see that this is equivalent to the curve
\eqref{eq:IR-curve} of the $(A_1,D_{2n-1})$ theory. Therefore, the
relevant deformation from $(A_1,D_{2n})$ to $(A_1,D_{2n-1})$ is
interpreted on the 2D side as $\Lambda_{2n}\to 0$.

This interpretation of the 4D RG-flow can also be verified
by the identification  \eqref{eq:parameters-4D2D} of the 4D parameters in terms of the 2D
parameters. Indeed, taking $\Lambda_{2n} \to
0$ with the other $\Lambda_i$ kept fixed leads to the limit $d_1 \to \infty$ with
$(d_1)^{-\frac{2n-i}{2n-1}}d_i$ kept fixed for $i\geq 2$, which is
precisely the limit we have taken to derive the IR curve
\eqref{eq:IR-curve} in the previous sub-section.

Note that the above limit is equivalent to taking 
\begin{align}
 c_n \to 0~,
\end{align}
with
\begin{align}
 \Lambda_k = -\sum_{i=k-n}^n c_ic_{k-i} + \delta_{k,n} (n+1)Qc_n \quad
 \text{for}\quad i=n,\cdots,2n-1~,
\label{eq:finite_lambda}
\end{align}
kept fixed. Note that $c_0,\cdots,c_{n-1}$ must be divergent to keep
\eqref{eq:finite_lambda} finite in this limit.

Given the above interpretation, it is now straightforward to take the
limit from $|I^{(n)}\rangle$ to
$|I^{(n-\frac{1}{2})}\rangle$ for an arbitrary integer $n$. Indeed, all we need to do is to take the
$c_n\to 0$ limit of the differential equations \eqref{eq:diff-eq-int}
with \eqref{eq:finite_lambda} kept finite. One obstacle to overcome here
is that some of the terms in the differential equations
\eqref{eq:diff-eq-int} will be divergent in this limit. This is because
$c_0,\cdots, c_{n-2}$ and $c_{n-1}$ are all divergent to keep
\eqref{eq:finite_lambda} finite in the limit $c_n\to 0$. Our strategy is
then to
renormalize the irregular state $|I^{(n)}\rangle$ by some prefactor $\mathcal{N}(c_0,\cdots,c_{n})$
and take the limit $c_n\to 0$ with \eqref{eq:finite_lambda} kept fixed, i.e.,
\begin{align}
 |I^{(n-\frac{1}{2})}\rangle = \lim_{c_n\to
 0}\bigg(\mathcal{N}(c_0,\cdots,c_n)|I^{(n)}\rangle\bigg)~.
\label{eq:ren}
\end{align}
Here the
prefactor $\mathcal{N}(c_0,\cdots,c_n)$ is determined so that the differential
equations for the renormalized state $\mathcal{N}(c_0,\cdots,c_n)|I^{(n)}\rangle$ have no divergent
terms in the limit $c_n\to 0$. While it is not obvious to us whether such a
prefactor always exists for an arbitrary positive integer $n$, we have checked
that such a prefactor does exist for $n=1,2,3,\dots,7$. Below, we
demonstrate it for $n=2$ and $n=3$, and explicitly show the limit from
$|I^{(2)}\rangle$ to $|I^{(3/2)}\rangle$ and that from $|I^{(3)}\rangle$
to $|I^{(5/2)}\rangle$.

\subsection{Rank $3/2$ from rank two} 
\label{sec:3/2}

We here study the limit from the irregular state of rank two,
$|I^{(2)}\rangle$, to the irregular state of rank $3/2$,
$|I^{(3/2)}\rangle$. 
Recall that the rank-two irregular state satisfies the following
differential equations:
\begin{align}
	L_4\ket{I^{(2)}}&=-c_2^2\ket{I^{(2)}}~,\label{rank2-l4}\\
	L_3\ket{I^{(2)}}&=-2c_1c_2\ket{I^{(2)}}~,\label{rank2-l3}\\
	L_2\ket{I^{(2)}}&=\qty(-2c_0c_2-c_1^2+3Qc_2)\ket{I^{(2)}}~,\label{rank2-l2}\\
	L_1\ket{I^{(2)}}&=\qty(2c_1(Q-c_0)+c_2\pdv{c_1})\ket{I^{(2)}}~,\label{rank2-l1}\\
	L_0\ket{I^{(2)}}&=\qty(
c_0(Q-c_0)
+2c_2\pdv{c_2}+c_1\pdv{c_1})\ket{I^{(2)}}~,\label{rank2-l0}
\end{align}
and $L_{n>4}\ket{I^{(2)}}=0$. 
We take the limit $c_2 \to 0$ with 
\begin{align}
 \Lambda_3 = -2c_1c_2~,\qquad \Lambda_2 = -2c_0c_2 -c_1^2 + 3Qc_2~,
\end{align}
kept fixed. In terms of $c_2,\Lambda_3$ and $\Lambda_2$, the above
differential equations are written as
\begin{align}
 	L_4\ket{I^{(2)}}&=-c_2^2\ket{I^{(2)}}~,
\\
	L_3\ket{I^{(2)}}&=\Lambda_3\ket{I^{(2)}}~,
\\
	L_2\ket{I^{(2)}}&=\Lambda_2\ket{I^{(2)}}~,
\\
	L_1\ket{I^{(2)}}&=\left(\frac{\Lambda_3}{2c_2}Q-\frac{\Lambda_2\Lambda_3}{2c_2^2}-\frac{(\Lambda_3)^3}{8c_2^4}
+\Lambda_3\frac{\partial}{\partial\Lambda_2}
- c_2^2\frac{\partial}{\partial \Lambda_3}\right)\ket{I^{(2)}}~,
\label{eq:L1-rank2}
\\[2mm]
	L_0\ket{I^{(2)}}&=\Bigg(-\frac{
\big(4c_2^3Q-4c_2^2\Lambda_2-(\Lambda_3)^2)
\big(12c_2^3Q-4c_2^2\Lambda_2-(\Lambda_3\big)^2)}{64c_2^6}
\nonumber\\[-1mm]
&\qquad \qquad \qquad +
 3\Lambda_3\frac{\partial}{\partial \Lambda_3} +
 2\Lambda_2\frac{\partial}{\partial
 \Lambda_2}+2c_2\frac{\partial}{\partial c_2} \Bigg)\ket{I^{(2)}}~.
\label{eq:L0-rank2}
\end{align}

Clearly, if we defined $|I^{(3/2)}\rangle$ by $\lim_{c_2\to
0}|I^{(2)}\rangle$ with $\Lambda_2$ and $\Lambda_3$ kept fixed, the resulting differential equations would contain
divergent terms and not make sense. Instead, we renormalize
$|I^{(2)}\rangle$
and then take the $c_2\to 0$ limit as
\begin{align}
 |I^{(3/2)}\rangle \equiv \lim_{c_2\to 0}
 \bigg(\mathcal{N}(c_0,c_1,c_2)|I^{(2)}\rangle\bigg)~,
\label{eq:ren-rank2}
\end{align}
where 
\begin{align}
\mathcal{N}(c_0,c_1,c_2)=
 c_2^{\frac{1}{2}c_0(Q-c_0)}\exp\left(-(c_0-Q)\frac{c_1^2}{c_2}\right)~.
\label{eq:prefactor1}
\end{align}
We stress here that $\Lambda_2$ and $\Lambda_3$ are kept fixed in the limit
$c_2\to 0$ in \eqref{eq:ren-rank2}.
We see that the prefactor \eqref{eq:prefactor1} removes all the
divergent terms in the equations \eqref{eq:L1-rank2} and
\eqref{eq:L0-rank2}, and as a result, we obtain the following
differential equations 
for $|I^{(3/2)}\rangle$:
\begin{align}
	L_3\ket{I^{(3/2)}}&=\Lambda_3\ket{I^{(3/2)}}~,
\label{eq:L3-3/2}
\\
	L_2\ket{I^{(3/2)}}&=\Lambda_2\ket{I^{(3/2)}}~,
\\
	L_1\ket{I^{(3/2)}}&=\Lambda_3\frac{\partial}{\partial\Lambda_2}\ket{I^{(3/2)}}~,
\\
	L_0\ket{I^{(3/2)}}&=\left(
 3\Lambda_3\frac{\partial}{\partial \Lambda_3} +
 2\Lambda_2\frac{\partial}{\partial
 \Lambda_2} \right)\ket{I^{(3/2)}}~,
\label{eq:L0-3/2}
\end{align}
and $L_{n>3}|I^{(3/2)}\rangle = 0$.

Note that \eqref{eq:prefactor1} is not the unique possible
renormalization factor. Different renormalization factors lead to
different expressions for the differential equations that are related to
each other by finite renormalization. For instance, instead of
\eqref{eq:prefactor1}, one can use
\begin{align}
 \mathcal{N}(c_0,c_1,c_2) =
c_1^{\frac{(2c_0-3Q)^2}{4}}
 c_2^{-c_0^2+2Qc_0-\frac{9Q^2}{8}}\exp\left(\frac{c_1^4}{16c_2^2} -
 \frac{(2c_0-Q)c_1^2}{4c_2} \right)~,
\end{align}
which leads to
\begin{align}
 L_3|I^{(3/2)}\rangle &= \Lambda_3|I^{(3/2)}\rangle~,
\label{eq:L3-3/2-v2}
\\
L_2 |I^{(3/2)}\rangle &= \Lambda_2|I^{(3/2)}\rangle~,
\\
L_1 |I^{(3/2)}\rangle &= \left(\frac{(\Lambda_2)^2}{2\Lambda_3}+
 \Lambda_3\frac{\partial}{\partial \Lambda_2}\right)|I^{(3/2)}\rangle~,
\\
L_0 |I^{(3/2)}\rangle &= \left(3\Lambda_3\frac{\partial}{\partial
 \Lambda_3} + 2\Lambda_2\frac{\partial}{\partial
 \Lambda_2}\right)|I^{(3/2)}\rangle~.
\label{eq:L0-3/2-v2}
\end{align}
The resulting equations \eqref{eq:L3-3/2-v2} -- \eqref{eq:L0-3/2-v2} are
obtained from the previous ones \eqref{eq:L3-3/2} -- \eqref{eq:L0-3/2} by
the following finite renormalization
\begin{align}
 |I^{(3/2)}\rangle \to \exp\left(\frac{(\Lambda_2)^3}{6(\Lambda_3)^2}\right)|I^{(3/2)}\rangle~.
\end{align}

\subsection{Rank $5/2$ from rank three}

While the differential equations for $|I^{(3/2)}\rangle$ obtained in the
previous sub-section is quite simple,
equations for $|I^{(n-\frac{1}{2})}\rangle$ for a
larger integer $n$ is more complicated. To demonstrate it, we here study
a similar limit from $|I^{(3)}\rangle$ to $|I^{(5/2)}\rangle$.

We start with the differential equations for the rank-three state:
\begin{align}
 L_6|I^{(3)}\rangle &= -c_3^2|I^{(3)}\rangle~,
\\
L_5|I^{(3)}\rangle &= -2c_3c_2|I^{(3)}\rangle~,
\\
L_4|I^{(3)}\rangle &= -\left(c_2^2+2c_3c_1\right)|I^{(3)}\rangle~,
\\
L_3|I^{(3)}\rangle &= -\left(2c_1c_2+2c_3(c_0-2Q)\right)|I^{(3)}\rangle~,
\\
L_2|I^{(3)}\rangle &= \left( -
 c_2(2c_0-3Q)-c_1^2 + c_3\frac{\partial}{\partial c_1}\right) |I^{(3)}\rangle~,
\label{eq:L2-rank3}
\\
L_1|I^{(3)}\rangle &= \left(
 -2c_1(c_0-Q)+2c_3\frac{\partial}{\partial c_2} +
 c_2\frac{\partial}{\partial c_1}\right)|I^{(3)}\rangle~,
\\
L_0|I^{(3)}\rangle &= \left(c_0(Q-c_0)+3c_3\frac{\partial}{\partial c_3}
 + 2 c_2\frac{\partial}{\partial c_2} + c_1 \frac{\partial}{\partial
 c_1}\right)|I^{(3)}\rangle~,
\label{eq:L0-rank3}
\end{align}
and $L_{n>6}|I^{(3)}\rangle = 0$. We take the limit $c_3\to 0$ with
\begin{align}
 \Lambda_5 = -2c_3c_2~,\qquad \Lambda_4 =
 -\left(c_2^2+2c_3c_1\right)~,\qquad \Lambda_3 = -\left(2c_1c_2 +
 2c_3(c_0 -2Q)\right)~,
\end{align}
kept fixed. Since this limit makes $c_0,c_1$ and $c_2$ divergent, so are
the
terms without derivatives
in equations \eqref{eq:L2-rank3} -- \eqref{eq:L0-rank3}. To remove these
divergent terms,
we renormalize the irregular state $|I^{(3)}\rangle$ so that
\begin{align}
 |I^{(5/2)}\rangle \equiv \lim_{c_3\to 0}\bigg(\mathcal{N}(c_0,c_1,c_2,c_3)|I^{(3)}\rangle\bigg)~.
\end{align}
where
\begin{align}
 \mathcal{N}(c_0,c_1,c_2,c_3) &= c_2^{(c_0-2Q)^2}c_3^{-(c_0-\frac{3}{2}Q)^2-\frac{5Q^2}{12}}\exp\Bigg(\frac{c_2^6}{96c_3^4}
 -\frac{c_1c_2^4}{12c_3^3}+\frac{c_2^2(3c_1^2+c_0c_2)}{12c_3^2}
\nonumber\\
&\qquad \qquad \qquad  -
 \frac{(c_0-Q)c_1c_2}{c_3}-\frac{c_1^4}{6c_2^2}+\frac{(c_0-2Q)c_1^2}{c_2}\Bigg)~.
\end{align}
Note that $\Lambda_3,\Lambda_4$ and $\Lambda_5$ are kept fixed in the
limit $c_3\to 0$.
 With the above definition, we find that $|I^{(5/2)}\rangle$ satisfies
\begin{align}
 L_5|I^{(5/2)}\rangle &= \Lambda_5|I^{(5/2)}\rangle~,
\label{eq:L5-5/2}
\\
L_4|I^{(5/2)}\rangle &= \Lambda_4|I^{(5/2)}\rangle~,
\\
L_3|I^{(5/2)}\rangle &= \Lambda_3|I^{(5/2)}\rangle~,
\\
L_2|I^{(5/2)}\rangle &= \left(
 -\frac{(\Lambda_4)^3}{3(\Lambda_5)^2}+\frac{\Lambda_3\Lambda_4}{\Lambda_5}   +\Lambda_5\frac{\partial}{\partial
 \Lambda_3} \right)|I^{(5/2)}\rangle~,
\label{eq:L2-5/2}
\\
L_1|I^{(5/2)}\rangle &= \left(\frac{(\Lambda_4)^4}{3(\Lambda_5)^3}-\frac{\Lambda_3(\Lambda_4)^2}{(\Lambda_5)^2}+\frac{(\Lambda_3)^2}{\Lambda_5} +
 3\Lambda_5\frac{\partial}{\partial \Lambda_4} +
 2\Lambda_4\frac{\partial}{\partial \Lambda_3}\right)|I^{(5/2)}\rangle~,
\label{eq:L1-5/2}
\\
L_0|I^{(5/2)}\rangle &= \left(5\Lambda_5\frac{\partial}{\partial
 \Lambda_5} + 4\Lambda_4\frac{\partial}{\partial \Lambda_4} +
 3\Lambda_3\frac{\partial}{\partial \Lambda_3}\right)|I^{(5/2)}\rangle~,
\label{eq:L0-5/2}
\end{align}
and $L_{k>6}|I^{(5/2)}\rangle = 0$. Note that \eqref{eq:L2-5/2} and
\eqref{eq:L1-5/2} contain 
terms that are divergent in the limit $\Lambda_5 \to 0$. One can remove
some of these terms by choosing a different renormalization factor
$\mathcal{N}(c_0,c_1,c_2,c_3)$, but unlike the rank-$3/2$ case, it is
not possible to remove all these terms. Indeed, $[L_2,L_1]|I^{(5/2)}\rangle = L_3|I^{(5/2)}\rangle$
implies that one cannot get rid of all these terms.

Note that the above differential equations \eqref{eq:L5-5/2} --
\eqref{eq:L0-5/2} are equivalent to the equations obtained in
\cite{Poghosyan:2023zvy}. Indeed, by performing the finite renormalization
\begin{align}
 |I^{(5/2)}\rangle \to
 \exp\left( -\frac{(\Lambda_4)^5}{15(\Lambda_5)^4} +
 \frac{\Lambda_3 (\Lambda_4)^3}{3(\Lambda_5)^3}-\frac{(\Lambda_3)^2\Lambda_4}{2(\Lambda_5)^2}\right)|I^{(5/2)}\rangle~,
\end{align}
and then redefining parameters as
\begin{align}
 \Lambda_5 \to -\Lambda_5~,\qquad \Lambda_4 \to -c_2^2~,\qquad \Lambda_3
 \to -2c_1c_2~,
\end{align}
our \eqref{eq:L5-5/2} --
\eqref{eq:L0-5/2} turn out to be mapped to Eqs.~(2.7) and (2.8) of
\cite{Poghosyan:2023zvy}. Thus, we have shown that the differential equations
discovered in \cite{Poghosyan:2023zvy} can be reproduced as a limit of the
differential equations for $|I^{(3)}\rangle$. As we have seen already, this limit
is interpreted as a 2D counterpart of a 4D RG-flow between two AD theories.

\section{Irregular states of general half-integer ranks}
\label{sec:formula}

We have seen in the previous section that the irregular states
$|I^{(3/2)}\rangle$  and $|I^{(5/2)}\rangle$ can be obtained as a
limit of (renormalized) $|I^{(2)}\rangle$ and $|I^{(3)}\rangle$,
respectively. We have checked that a similar limit from
$|I^{(n)}\rangle$ to
$|I^{(n-\frac{1}{2})}\rangle$ exists for $n=1,2,3,\cdots,7$. All these limits
are interpreted as the 2D counterpart of 4D RG-flow from
$(A_1,D_{2n})$ to $(A_1,D_{2n-1})$.
Using this limit,
one can read off differential equations satisfied by
$|I^{(n-\frac{1}{2})}\rangle$ for these values of $n$.
Instead of writing down these equations separately, we here summarize
them into a general formula, and then conjecture that it holds for all
irregular states of half-integer ranks.

\subsection{Differential equation for general $|I^{(n-\frac{1}{2})}\rangle$}

Our conjecture for the differential equations for
$|I^{(n-\frac{1}{2})}\rangle$ for a general positive integer $n$ is the
following:
\begin{align}
 L_k|I^{(n-\frac{1}{2})}\rangle = \left\{
\begin{array}{l}
0 \qquad (2n\leq k)
\\[3mm]
\Lambda_k|I^{(n-\frac{1}{2})}\rangle\qquad (n\leq k\leq 2n-1)
\\[3mm]
{\displaystyle \bigg(f_k(\Lambda_n,\Lambda_{n+1},\cdots,\Lambda_{2n-1}) +
 \sum_{m=n}^{2n-k-1}(m-k)\Lambda_{m+k}\frac{\partial}{\partial \Lambda_m}\bigg)|I^{(n-\frac{1}{2})}\rangle\qquad (0\leq k\leq n-1)}
\end{array}
\right.~.
\label{eq:diff-eq-halfint}
\end{align}
Here, the functions $f_0,\cdots,f_{n-1}$ are defined iteratively by 
\begin{align}
	\label{ourrep}
 f_{k}(\Lambda) 
&=  \sum_{\ell=1}^{2n-k-2}(-1)^{\ell-1}\frac{(\ell+n-k-1)
 !}{(n-k-1)!}\,\frac{g_{\ell+1}^{(2n-1)\ell+k}(\Lambda)}{(\Lambda_{2n-1})^\ell} 
\nonumber\\
&\qquad + \sum_{m=1}^{n-k-1}
 \left(\sum_{\ell=1}^{m}(-1)^{\ell-1}\frac{(\ell+n-k-1)!}{(n-k-1)!}\frac{g_{\ell}^{(2n-1)\ell-m}(\Lambda)}{(\Lambda_{2n-1})^\ell}\right)
 f_{k+m}(\Lambda)
\end{align}
with
\begin{align}
 g_{m}^i(\Lambda) \equiv \frac{1}{2\pi
 i}\oint_{|z|=1}\frac{dz}{z^{1+i}}\frac{1}{m!}\left(\sum_{\ell=n}^{2n-2}z^\ell\Lambda_\ell\right)^{m}~.
\label{eq:g}
\end{align}

Note that $f_k$ and
$g_m^i$ implicitly depend on $n$. Note also that \eqref{ourrep} always implies that
\begin{align}
 f_{n-1}(\Lambda) = \sum_{\ell=1}^{n-1}(-1)^{\ell-1} \ell!
 \frac{g_{\ell+1}^{(2n-1)\ell+n-1}(\Lambda)}{(\Lambda_{2n-1})^\ell}~,
\label{eq:fn-1}
\end{align}
which then determines the other $f_k(\Lambda)$ iteratively through
\eqref{ourrep}. The function $g_m^i(\Lambda)$ defined by \eqref{eq:g} is a polynomial of
$\Lambda_{n},\cdots,\Lambda_{2n-2}$ with a special property. To see
this, let us
assign {\it weight} $k$ to $\Lambda_k$ for $k=n,\cdots,2n-1$. Then
it turns out that $g_m^i(\Lambda)$ is a linear combination of
degree-$m$ monomials of weight $i$. For instance, when $n=4$, we find
\begin{align}
 g_1^4(\Lambda) = \Lambda_4~,\quad g_2^8(\Lambda) =
 \frac{(\Lambda_4)^2}{2!}~,\quad g_2^9(\Lambda) = \Lambda_4
 \Lambda_5~,\quad g_2^{10}(\Lambda) = \Lambda_4\Lambda_6 +
 \frac{(\Lambda_5)^2}{2!}~,\quad \cdots~.
\end{align}
This fact, combined with \eqref{eq:fn-1} and \eqref{ourrep}, then
implies that $f_k(\Lambda)$ is a weight-$k$ Laurent polynomial of
$\Lambda_n,\cdots,\Lambda_{2n-1}$, and  vanishes in the limit $\Lambda_{2n-1}\to
\infty$ with $\Lambda_n,\cdots,\Lambda_{2n-2}$ kept fixed. Explicit
expressions for $f_k(\Lambda)$ for some lower values of $n$ are shown in
Appendix \ref{app:f}.

While we have no general proof of our conjecture \eqref{eq:diff-eq-halfint}, we have checked for $n=1,2,3,\cdots,7$ that there exists a limit 
\begin{align}
 |I^{(n-\frac{1}{2})}\rangle = \lim_{c_n\to 0}\mathcal{N}(c_0,\cdots,c_n)|I^{(n)}\rangle~,
\end{align}
which maps
the differential equations \eqref{eq:diff-eq-int} for $|I^{(n)}\rangle$
to our conjectured equations \eqref{eq:diff-eq-halfint} for
$|I^{(n-\frac{1}{2})}\rangle$.\footnote{We used
Mathematica for this computation. For $n=7$, it took our code of
Mathematica running on a Mac mini about seventy minutes to
find the prefactor $\mathcal{N}(c_0,\cdots,c_n)$.} This is a strong evidence for our
conjecture. For instance, \eqref{eq:L3-3/2-v2} -- \eqref{eq:L0-3/2-v2}
for $|I^{(3/2)}\rangle$ and \eqref{eq:L5-5/2} -- \eqref{eq:L0-5/2} for
$|I^{(5/2)}\rangle$ are precisely reproduced from our general conjecture
\eqref{eq:diff-eq-halfint}.

\subsection{Representation of Virasoro algebra}

Our conjecture \eqref{eq:diff-eq-halfint} is expected to give a
representation of the Virasoro algebra. Indeed, we have checked for $n=1,2,\cdots,11$ that
\eqref{eq:diff-eq-halfint} is consistent with the Virasoro, i.e.,
\begin{align}
 L_kL_m|I^{(n-\frac{1}{2})}\rangle - L_mL_k|I^{(n-\frac{1}{2})}\rangle =
 (k-m)L_{k+m}|I^{(n-\frac{1}{2})}\rangle~,
\label{eq:Vir+}
\end{align}
for $k,m \geq 0$.
It is left for future work to prove that \eqref{eq:diff-eq-halfint} is
consistent with 
\eqref{eq:Vir+}
for a general integer $n$.

In the rest of this sub-section, we show one interesting fact on our
conjectured formula \eqref{eq:diff-eq-halfint}; {\it assuming
\eqref{eq:diff-eq-halfint} correctly gives a representation of the
Virasoro algebra, one can prove that}
\begin{align}
 f_0(\Lambda) = 0~,
\label{eq:f0}
\end{align}
{\it for all positive integers $n$.} The proof goes as follows. When
\eqref{eq:diff-eq-halfint} gives a representation of the Virasoro
algebra, it follows that
\begin{align}
 [L_k,\, L_0]|I^{(n-\frac{1}{2})}\rangle = k\,L_{k}|I^{(n-\frac{1}{2})}\rangle~.
\end{align}
This implies that the functions $f_k$ in \eqref{eq:diff-eq-halfint} must
satisfy
\begin{align}
\sum_{m=n}^{2n-1}m\,\Lambda_{m}\frac{\partial f_k}{\partial \Lambda_m} -
 \sum_{m=n}^{2n-k-1}(m-k)\Lambda_{m+k}\frac{\partial f_0}{\partial
 \Lambda_{m}} =
 kf_k~.
\label{eq:f_must_satisfy}
\end{align}
As we have seen, $f_k(\Lambda)$ is a
weight-$k$ Laurent polynomial of $\Lambda_n,\cdots,\Lambda_{2n-1}$, and therefore
$\left(\sum_{m=n}^{2n-1} m\,\Lambda_m\frac{\partial}{\partial \Lambda_m}\right)f_k(\Lambda) = k\,f_k(\Lambda)$.
This means that \eqref{eq:f_must_satisfy} is equivalent to the condition that
\begin{align}
 A_k\equiv  \sum_{m=n}^{2n-k-1}(m-k)\Lambda_{m+k}\frac{\partial f_0}{\partial
 \Lambda_{m}}
\end{align}
vanishes for $k=0,1,\cdots,n-1$. Now, this condition is further identical to
the condition that
\begin{align}
 \frac{\partial f_0}{\partial \Lambda_m} = 0 
\end{align}
for $m=n,\cdots, 2n-1$, since $\partial f_0/\partial \Lambda_m$ can be
written as a linear combination of $A_0,\cdots,A_{n-1}$. Thus,
we see that $f_0(\Lambda)$ is independent of all
$\Lambda_n,\Lambda_{n+1},\cdots,\Lambda_{2n-2}$ and
$\Lambda_{2n-1}$. This and the fact that $f_0(\Lambda)\to 0$ in the
limit $\Lambda_{2n-1}\to \infty$ imply \eqref{eq:f0}.

\subsection{Partition function and parameters of $(A_1,D_{2n-1})$}

The AGT correspondence suggests that the Nekrasov partition function of
the $(A_1,D_{2n-1})$ theory is written as
\begin{align}
 \mathcal{Z}_{(A_1,D_{2n-1})} = \langle \Delta
 |I^{(n-\frac{1}{2})}\rangle~.
\label{eq:part-D2n-1}
\end{align}
Note here that, since the differential equations \eqref{eq:diff-eq-halfint}
do not fix the overall normalization constant of
$|I^{(n-\frac{1}{2})}\rangle$, the right-hand side can be
computed only up to a constant prefactor. However, this ambiguous
prefactor is independent of $\Lambda_n,\cdots,\Lambda_{2n-2}$ and
$\Lambda_{2n-1}$ because changing the
$\Lambda_i$-dependence of $|I^{(n-\frac{1}{2})}\rangle$ spoils the
differential equations \eqref{eq:diff-eq-halfint}.
Therefore, the dependence of \eqref{eq:part-D2n-1}
on $\Lambda_n,\cdots,\Lambda_{2n-2}$ and
$\Lambda_{2n-1}$ is completely fixed by the differential equations
\eqref{eq:diff-eq-halfint}.

Below, we show that all the 4D parameters of the $(A_1,D_{2n-1})$
theory are also fixed by the differential equations
\eqref{eq:diff-eq-halfint} and therefore independent of the ambiguous
normalization constant of $|I^{(n-\frac{1}{2})}\rangle$.
To see this, let us identify the precise relation between the 4D and 2D
parameters. Note first that the
SW curve of the $(A_1,D_{2n-1})$ theory is identified as
\begin{align}
 x^2 = -\frac{\langle \Delta |T(z)|I^{(n-\frac{1}{2})}\rangle}{\langle
 \Delta |I^{(n-\frac{1}{2})}\rangle}~.
\end{align}
The right-hand side can be evaluated by using \eqref{eq:diff-eq-halfint} as
\begin{align}
 x^2 = -\Lambda_{2n-1}z^{2n-3} - \sum_{i=1}^{n-1} \Lambda_{2n-1-i}z^{2n-3-i} - \sum_{i=1}^{n-1}v_iz^{n-2-i} -
 \frac{\Delta}{z^2}~,
\label{eq:curve-35}
\end{align}
where $v_i$ is defined by
\begin{align}
 v_i = \frac{\langle \Delta |L_{n-i}|I^{(n-\frac{1}{2})}\rangle}{\langle
 \Delta |I^{(n-\frac{1}{2})}\rangle}~.
\label{eq:vi-2}
\end{align}
Here, the SW 1-form is given by $\lambda \equiv xdz$. 
By a $\lambda$-preserving change of coordinates, the curve \eqref{eq:curve-35} is rewritten as
\begin{align}
 x^2 = z^{2n-3} - \sum_{i=1}^{n-1} \frac{\Lambda_{2n-1-i}}{(-\Lambda_{2n-1})^{\frac{2n-1-i}{2n-1}}}z^{2n-3-i} - \sum_{i=1}^{n-1}\frac{v_i}{(-\Lambda_{2n-1})^{\frac{n-i}{2n-1}}}z^{n-2-i} -
 \frac{\Delta}{z^2}~,
\label{eq:curve-3}
\end{align}
Comparing \eqref{eq:curve-3} with \eqref{eq:IR-curve}, we find
the following identification between the 4D and 2D parameters:
\begin{align}
 \tilde{d}_i =
 -\frac{\Lambda_{2n-1-i}}{(-\Lambda_{2n-1})^{\frac{2n-1-i}{2n-1}}}~,\qquad
 \tilde{u}_i = -\frac{v_i}{(-\Lambda_{2n-1})^{\frac{n-i}{2n-1}}}~,\qquad
 m = \sqrt{-\Delta }~.
\label{eq:param-id-halfint}
\end{align}
Note that the expression \eqref{eq:vi-2} for $(A_1,D_{2n-1})$ is regarded as the $c_n\to 0$
limit of \eqref{eq:vi} for $(A_1,D_{2n})$, since our
$|I^{(n-\frac{1}{2})}\rangle$ is expected to be obtained by \eqref{eq:ren}.

From \eqref{eq:param-id-halfint}, we see that the relevant couplings
$\tilde{d}_i$ are completely fixed by
$\Lambda_n,\cdots,\Lambda_{2n-1}$, and the mass parameter $m$ is fixed
by $\Delta$. Therefore, $\tilde{d}_i$ and $m$ are independent of the
ambiguous normalization constant of $|I^{(n-\frac{1}{2})}\rangle$.
Similarly, from \eqref{eq:vi-2} and
\eqref{eq:param-id-halfint}, we see that the VEVs of the Coulomb branch
parameters $\tilde{u_i}$ are all fixed in terms of the ratio of
$\partial /\partial \Lambda_i \langle \Delta|I^{(n-\frac{1}{2})}\rangle$
and $\langle \Delta |I^{(n-\frac{1}{2})}\rangle$, and therefore independent of the
ambiguous normalization constant of $|I^{(n-\frac{1}{2})}\rangle$.

In the next section, we will demonstrate for $n=2$ how to evaluate the partition
function \eqref{eq:part-D2n-1} by solving the differential equations \eqref{eq:diff-eq-halfint}.

\section{Solution to the differential equations for $|I^{(3/2)}\rangle$}
\label{sec:example}

In this section, we 
study our conjectured differential
equations \eqref{eq:diff-eq-halfint}.
The basic strategy to solve these equations is the same as those in
\cite{Gaiotto:2012sf, Nishinaka:2019nuy, Poghosyan:2023zvy}; we assume
that $|I^{(n-\frac{1}{2})}\rangle$ is expanded in terms of
generalized descendants of a lower-rank irregular state. Here, we mean by
a generalized descendant a linear combination of states obtained by
acting $L_{k}$ for $k<0$ and/or $\frac{\partial}{\partial \Lambda_\ell}$ for $n\leq \ell
\leq 2n-2$ on a lower-rank irregular state.\footnote{The remaining eigenvalue
$\Lambda_{2n-1}$ will be an expansion parameter instead.}

One question here is that of which lower-rank state we should consider
a generalized descendant. For the integer-rank irregular state
$|I^{(n)}\rangle$ satisfying \eqref{eq:diff-eq-int}, it was conjectured
in \cite{Gaiotto:2012sf} that there exists a series expansion of the form
\begin{align}
 |I^{(n)}\rangle = \mathcal{M}(c_0,\cdots,c_n)\sum_{k=0}^\infty (c_n)^k
 |I_k^{(n-1)}\rangle~,
\label{eq:ansatz_n}
\end{align}
where $|I^{(n-1)}_k\rangle$ are generalized descendants of
$|I^{(n-1)}\rangle$ such that $|I^{(n-1)}_0\rangle =
|I^{(n-1)}\rangle$, and $\mathcal{M}(c_0,\cdots,c_n)$ is some function
of $c_0,\cdots, c_{n}$ that makes it possible to solve
\eqref{eq:diff-eq-int} order by order in $c_n$.\footnote{The prefactor
$\mathcal{M}(c_0,\cdots,c_n)$ also depends on some other parameter than $c_0,\cdots, c_n$.} Similarly, for the rank $5/2$ state $|I^{(5/2)}\rangle$, the
authors of \cite{Poghosyan:2023zvy} found a series expansion of the form
\begin{align}
|I^{(5/2)}\rangle =
 \mathcal{M}(\Lambda_3,\Lambda_4,\Lambda_5)\sum_{k=0}^\infty (\Lambda_5)^k|I^{(2)}_k\rangle~,
\end{align}
where $|I^{(2)}_k\rangle$ are generalized descendants of
$|I^{(2)}\rangle$ such that $|I^{(2)}_0\rangle =
|I^{(2)}\rangle$.\footnote{As in the previous equation, the prefactor
$\mathcal{M}(\Lambda_3,\Lambda_4,\Lambda_5)$ also depends on an extra parameter.} Given
the above success of the earlier works, we conjecture that the solution to our differential
equations \eqref{eq:diff-eq-halfint} has the following series expansion
\begin{align}
 |I^{(n-\frac{1}{2})}\rangle =
 \mathcal{M}(\Lambda_n,\cdots,\Lambda_{2n-1})\sum_{k=0}^\infty
 (\Lambda_{2n-1})^{k}|I^{(n-1)}_k\rangle~,
\label{eq:ansatz_n-1/2}
\end{align}
where $|I^{(n-1)}_k\rangle$ are generalized descendants of
$|I^{(n-1)}\rangle$ such that $|I^{(n-1)}_0\rangle = |I^{(n-1)}\rangle$, and $\mathcal{M}(\Lambda_n,\cdots,\Lambda_{2n-1})$
is a function of $\Lambda_n,\cdots,\Lambda_{2n-1}$ which enables us to
solve \eqref{eq:diff-eq-halfint} order by order in $\Lambda_{2n-1}$.
Below, we verify this conjecture for the rank $3/2$ state
$|I^{(3/2)}\rangle$.

After finding an expression for $|I^{(3/2)}\rangle$ of the form
\eqref{eq:ansatz_n-1/2} in Sec.~\ref{subsec:solve-3/2}, we will evaluate
the Nekrasov partition function of the $(A_1,D_3)$ theory, i.e.,
$\mathcal{Z}_{(A_1,D_3)} = \langle \Delta |I^{(3/2)}\rangle$ in
Sec.~\ref{subsec:Z_A1D3}.

\subsection{Solving rank $3/2$}
\label{subsec:solve-3/2}

We now focus on the case of rank $3/2$. While our general conjecture
\eqref{eq:diff-eq-halfint} gives \eqref{eq:L3-3/2-v2} --
\eqref{eq:L0-3/2-v2} for $|I^{(3/2)}\rangle$, we have already shown that these equations are
mapped to \eqref{eq:L3-3/2} -- \eqref{eq:L0-3/2} by a finite
renormalization of the irregular state $|I^{(3/2)}\rangle$.
Therefore, in this sub-section, we solve \eqref{eq:L3-3/2} --
\eqref{eq:L0-3/2} (together with $L_{m\geq 4}|I^{(3/2)}\rangle = 0$) and
find a power series expression for
$|I^{(3/2)}\rangle$.

To that end, we first redefine $\Lambda_2$ as
\begin{align}
 \Lambda_2 \equiv -c_1^2~,
\end{align}
and treat $c_1$ and $\Lambda_3$ as independent variables. Then the
differential equations are
expressed as
\begin{align}
L_{m\geq 4}|I^{(3/2)}\rangle &= 0~,
\\
 	L_3\ket{I^{(3/2)}}&=\Lambda_3\ket{I^{(3/2)}}~,
\label{eq:L3-3/2-v4}
\\
	L_2\ket{I^{(3/2)}}&=-c_1^2\ket{I^{(3/2)}}~,
\\
	L_1\ket{I^{(3/2)}}&=
 -\frac{\Lambda_3}{2c_1}\frac{\partial}{\partial c_1}\ket{I^{(3/2)}}~,
\\
	L_0\ket{I^{(3/2)}}&=\left(
 3\Lambda_3\frac{\partial}{\partial \Lambda_3} +
 c_1\frac{\partial}{\partial
 c_1} \right)\ket{I^{(3/2)}}~.
\label{eq:L0-3/2-v4}
\end{align}
Now, we consider the following ansatz for the solution to the above equations:
\begin{align}
	\ket{I^{(3/2)}}
	=(\Lambda_3)^{\frac{\beta_1(Q-\beta_1)}{3}}\cdot\qty(\frac{c_1^3}{\Lambda_3})^{\frac{\epsilon}{3}}\exp(-\frac{4(Q-\beta_1)}{3}\frac{c_1^3}{\Lambda_3})\sum_{k=0}^\infty(\Lambda_3)^k\ket{I_k^{(1)}}~,
	\label{ansatz3/2}
\end{align}
where  $|I^{(1)}_k\rangle$ are generalized descendants of
$|I^{(1)}\rangle$ satisfying
\begin{align}
L_{m\geq 3}|I^{(1)}\rangle &=0~,
\\
 L_2|I^{(1)}\rangle &= -c_1^2|I^{(1)}\rangle~,
\label{eq:L2-I1}
\\
L_1|I^{(1)}\rangle &= 2c_1(Q-\beta_1)|I^{(1)}\rangle~,
\\
L_0|I^{(1)}\rangle &= \left(\beta_1(Q-\beta_1)+
 c_1\frac{\partial}{\partial c_1} \right)|I^{(1)}\rangle~.
\label{eq:L0-I1}
\end{align}
The parameter $\epsilon$ in \eqref{ansatz3/2} will be fixed as a
function of $\beta_1$ and $Q$ below.

The prefactor of the $\Lambda_3$-expansion in \eqref{ansatz3/2} is chosen so that the differential
equations \eqref{eq:L3-3/2-v4} -- \eqref{eq:L0-3/2-v4} can be solved order by order in $\Lambda_3$.
Indeed, substituting \eqref{ansatz3/2} into \eqref{eq:L3-3/2-v4} --
\eqref{eq:L0-3/2-v4}, we obtain the following recursive equations for
the generalized descendants $|I^{(1)}_k\rangle$:
\begin{align}
 L_{m\geq 4}|I^{(1)}_k\rangle &= 0~,
\label{rank3/2rec-l4}
\\
	L_3\ket{I_k^{(1)}}&=\ket{I_{k-1}^{(1)}}~,\label{rank3/2rec-l3}\\
	L_2\ket{I_k^{(1)}}&=-c_1^2\ket{I_k^{(1)}}~,\label{rank3/2rec-l2}\\
	L_1\ket{I_k^{(1)}}&=2(Q-\beta_1) c_1\ket{I_k}+\qty(-\frac{\epsilon}{2c_1^2}-\frac{1}{2c_1}\pdv{c_1})\ket{I_{k-1}^{(1)}}~,\label{rank3/2rec-l1}\\
	L_0\ket{I_k^{(1)}}&=\qty(\beta_1(Q-\beta_1)+3k+c_1\pdv{c_1})\ket{I_k^{(1)}}~.\label{rank3/2rec-l0}
\end{align}
For $k=0$, these equations
reduce to 
\begin{align}
L_{m\geq 3}|I_0^{(1)}\rangle &= 0~,
\\
	L_2\ket{I_0^{(1)}}&=-c_1^2\ket{I_0^{(1)}}~,\label{rankone-l2}\\
	L_1\ket{I_0^{(1)}}&=2(Q-\beta_1) c_1\ket{I_0^{(1)}}~,\label{rankone-l1}\\
	L_0\ket{I_0^{(1)}}&=\qty(\beta_1(Q-\beta_1)+c_1\pdv{c_1})\ket{I_0^{(1)}}~,\label{rankone-l0}
\end{align}
which are identical to
\eqref{eq:L2-I1} -- \eqref{eq:L0-I1}. Therefore, we identify
\begin{align}
|I^{(1)}_0\rangle = |I^{(1)}\rangle~.
\label{eq:I1-0}
\end{align}
We see that the
 prefactor in \eqref{ansatz3/2} has been chosen so that \eqref{eq:I1-0}
 holds. The equations for $k\geq 1$ can then be solved recursively. For
 instance, from the equations for $k=1$, we find 
\begin{align}
 	\ket{I_1^{(1)}}
 = \qty(-\frac{1}{4(c_1)^2}L_{-1}-\frac{3(Q-\beta_1)}{4(c_1)^2}\pdv{c_1}+\frac{\nu_3}{c_1^3})\ket{I^{(1)}}~,
 \qquad \epsilon =3 Q^2 - 5 Q \beta_1 + 2\beta_1^2~.
\label{eq:I1}
\end{align}
The coefficient $\nu_3$ in the last term is not fixed at this order, but
will be fixed as 
\begin{align}
 \nu_3 = \frac{1}{24} (Q-\beta_1) \qty(-16 \beta_1^2-39Q^2+50\beta_1Q+1)~,
\end{align}
when considering the equations for $k=2$. Continuing this
recursive procedure, we obtain\footnote{We used Mathematica for this
recursive procedure. It took our code running on a Mac mini several minutes to find the expression
for $|I^{(1)}_3\rangle$.}
\begin{align}
 	\ket{I_2^{(1)}}&=\Bigg(
-\frac{5}{48c_1^4}L_{-2}+\frac{1}{32c_1^4}\qty(L_{-1})^2+\frac{39Q^3+44\beta_1-89Q^2\beta_1-16\beta_1^3+22Q(3\beta_1^2-2)}{96c_1^5}L_{-1}
\nonumber\\
&\qquad +\frac{3(Q-\beta_1)}{16c_1^4}L_{-1}\pdv{c_1}+\frac{-11+54Q^2-108Q\beta_1+54\beta_1^2}{192c_1^4}\pdv[2]{c_1}
\nonumber\\
&\qquad
 +\frac{1}{96c_1^5}\Big(3+117Q^4-384Q^3\beta_1-94\beta_1^2+48\beta_1^4+15Q^2(31\beta_1^2-8)
\nonumber\\
&\qquad \qquad \qquad +Q\beta_1(199
 -246\beta_1^2) \Big)\pdv{c_1} +\frac{\nu_4}{c_1^6}
	\Bigg)\ket{I^{(1)}}~,
\label{eq:I2}
\\
|I_3^{(1)}\rangle &= \Bigg(
-\frac{5}{96c_1^6}L_{-3}+\frac{5}{192c_1^6}L_{-2}L_{-1} -
 \frac{1}{384c_1^6}L_{-1}^3
 +\frac{5(Q-\beta_1)}{64c_1^6}L_{-2}\frac{\partial}{\partial
 c_1}-\frac{3(Q-\beta_1)}{128c_1^6}L_{-1}^2\frac{\partial}{\partial c_1}
\nonumber\\
&\qquad +
 \frac{5(Q-\beta_1)(16\beta_1^2-50Q\beta_1+39Q^2-91)}{1152c_1^7}L_{-2}
\nonumber\\
&\qquad
 - \frac{(Q-\beta_1)(16\beta_1^2-50Q\beta_1+39Q^2-87)}{768c_1^7}L_{-1}^2
\nonumber\\
&\qquad 
+\frac{1}{768c_1^6} \Big(-54 \beta _1^2-54 Q^2+108 \beta _1
 Q+11\Big)L_{-1}\frac{\partial^2}{\partial c_1^2}
\nonumber\\
&\qquad +\frac{1}{384c_1^7} \Big(-48 \beta
 _1^4+\beta _1^2 \left(223-465 Q^2\right)+\beta _1 Q \left(384
 Q^2-457\right)
\nonumber\\
&\qquad \qquad \qquad -3 \left(39 Q^4-83 Q^2+12\right)+246 \beta _1^3 Q\Big)L_{-1}\frac{\partial}{\partial c_1}
\nonumber\\
&\qquad +\frac{1}{4608c_1^8}\Big(-256 \beta _1^6+2112 Q\beta _1^5+\beta _1^4 \left(2176-7204 Q^2\right)+76 Q\beta _1^3 \left(171
 Q^2-154\right)
\nonumber\\
&\qquad \qquad \qquad +\beta _1^2 \left(-13069 Q^4+23372 Q^2-5907\right)+2 Q\beta _1 \left(3471 Q^4-10258 Q^2+6156\right)
\nonumber\\
&\qquad \qquad \qquad  -1521 Q^6+6672 Q^4-7485 Q^2+612\Big)L_{-1}
\nonumber\\
&\qquad -\frac{1}{256c_1^6} \left(Q-\beta _1\right) \left(18 Q^2+18
 \beta _1 \left(\beta _1-2 Q\right)-11\right)\frac{\partial^3}{\partial
 c_1^3}
\nonumber\\
&\qquad + \frac{1}{4608c_1^7}\Big(864 \beta _1^5-5292 Q\beta _1^4
 +2 \beta _1^3 \left(6399 Q^2-1753\right)-6Q \beta _1^2 \left(2547
 Q^2-1852\right)
\nonumber\\
&\qquad \qquad \qquad +\beta _1
 \left(9018 Q^4-12301 Q^2+1541\right)-2106 Q^5+4695 Q^3-1541
 Q\Big)\frac{\partial^2}{\partial c_1^2}
\nonumber\\
&\qquad -\frac{Q-\beta _1}{4608c_1^8}\Big(768 \beta _1^6-6336 Q\beta _1^5+4 \beta _1^4 \left(5403 Q^2-1328\right)-36 Q\beta _1^3
 \left(1083 Q^2-812\right)
\nonumber\\
&\qquad \qquad \qquad +\beta _1^2
 \left(39207 Q^4-60268 Q^2+11045\right)- Q\beta _1 \left(20826 Q^4-55278 Q^2+25766\right) 
\nonumber\\
&\qquad \qquad \qquad +3 \left(1521 Q^6-6360 Q^4+6907 Q^2-542\right)\Big)\frac{\partial}{\partial c_1}
\;  +\; \frac{\nu_5}{c_1^9}
\Bigg)|I^{(1)}\rangle~,
\label{eq:I3}
\end{align}
where
\begin{align}
 \nu_4 &=\frac{1}{1152} (Q-\beta_1) \Big(-256 \beta_1^5+800
 \beta_1^3-169 \beta_1+1521 Q^5-5421 \beta_1 Q^4+\left(7648
 \beta_1^2-3318\right) Q^3
\nonumber\\
	&\qquad\qquad +\left(6190 \beta_1-5348 \beta_1^3\right) Q^2+\left(1856
 \beta_1^4-3852 \beta_1^2+271\right) Q\Big)~,
\\
\nu_5 &= \frac{1}{82944}\Bigg(
4096 \beta_1 ^9 -50688 Q\beta _1^8 +192 \beta _1^7 \left(1445
 Q^2-196\right)+\beta _1^6 Q\left(374592-881352 Q^2\right)
\nonumber\\
&\qquad\qquad  + 12 \beta _1^5 \left(149197 Q^4-132397 Q^2+8932\right) -6 \beta
 _1^4 Q \left(401871 Q^4-619964 Q^2+134917\right)
\nonumber\\
&\qquad\qquad  +\beta _1^3 \left(2152493 Q^6-5189151 Q^4+2434023
 Q^2-48313\right)
\nonumber\\
&\qquad\qquad  + 3 \beta _1^2 Q \left(-409305 Q^6+1436827 Q^4-1211869
 Q^2+69691\right)
\nonumber\\
&\qquad\qquad +3 \beta _1 \left(135369 Q^8-657651 Q^6+897041 Q^4-98149
 Q^2+630\right)
\nonumber\\
&\qquad\qquad -Q(59319 Q^8-383643 Q^6+787221 Q^4-133687 Q^2+1890)
\Bigg)~.
\end{align}
Note that the central charge is given by $c=1+6Q^2$ in our convention for the Liouville charge.

\subsection{Partition function of $(A_1,D_3)$
}
\label{subsec:Z_A1D3}

Using the series expression for $|I^{(3/2)}\rangle$ obtained in the
previous sub-section, we now evaluate the
partition function of the $(A_1,D_3)$ theory $\mathcal{Z}_{(A_1,D_3)} =
\langle \Delta_{\beta_0} |I^{(3/2)}\rangle$.\footnote{$|\Delta_{\beta_0}\rangle$ is a Virasoro primary state with highest weight $\Delta_{\beta_0}\equiv\beta_0(Q-\beta_0)$.
}
From the expansion
\eqref{ansatz3/2}, we see that
\begin{align}
 \mathcal{Z}_{(A_1,D_3)} =
 (\Lambda_3)^{\frac{\beta_1(Q-\beta_1)}{3}}\left(\frac{(c_1)^3}{\Lambda_3}\right)^{\frac{\epsilon}{3}}\exp\left(-\frac{4(Q-\beta_1)}{3}\frac{(c_1)^3}{\Lambda_3}\right)\langle
 \Delta_{\beta_0}|I^{(1)}\rangle\sum_{k=0}^\infty
 \left(-\frac{3\Lambda_3}{4(c_1)^3}\right)^k D_k~,
\label{eq:ZA1D3}
\end{align}
where we used the shorthand
\begin{align}
 D_k \equiv \left(-\frac{4}{3}c_1^3\right)^k\frac{\langle \Delta_{\beta_0}|I_k^{(1)}\rangle}{\langle \Delta_{\beta_0}|I^{(1)}\rangle}~.
\end{align}
The factor $\langle
\Delta_{\beta_0}|I^{(1)}\rangle$ in \eqref{eq:ZA1D3} can be evaluated as
follows. From \eqref{eq:L0-I1}, it follows
that
\begin{align}
 \left(\beta_1(Q-\beta_1) -\beta_0(Q-\beta_0) + c_1\frac{\partial}{\partial
 c_1}\right)\langle \Delta_{\beta_0}|I^{(1)}\rangle &=0~,
\end{align}
which implies 
\begin{align}
 \langle \Delta_{\beta_0} |I^{(1)}\rangle =
c_1^{\beta_0(Q-\beta_0) -\beta_1(Q-\beta_1)}~,
\end{align}
 up to a constant prefactor. 
From \eqref{eq:I1}, \eqref{eq:I2} and \eqref{eq:I3}, we see that the
first few coefficients $D_0,D_1,D_2$ and $D_3$ are evaluated as
\begin{align}
D_0 &= 1~,
\\
 D_1 &= \frac{1}{18} \left(Q-\beta _1\right) \left(-18 \beta _0^2+39
 Q^2+18 \beta _0 Q+34 \beta _1 \left(\beta _1-2 Q\right)-1\right)~,
\label{eq:D1}
\\
D_2 &= \frac{1}{648} \bigg(6 \beta _0^4 \left(54(Q-\beta_1)^2-11\right)
-12 \beta _0^3 Q \left(54(Q-\beta_1)^2-11\right)
\nonumber\\
&\qquad -6 \beta _0^2 \left(-12 \beta _1 \left(2 Q-\beta _1\right)
 \left(32 Q^2+17 \beta _1 \left(\beta _1-2 Q\right)-22\right)+180
 Q^4-283 Q^2+17\right)
\nonumber\\
&\qquad +6 \beta _0 Q \left(-6 \beta _1 \left(2 Q-\beta _1\right)
 \left(73 Q^2+34 \beta _1 \left(\beta _1-2 Q\right)-44\right)+234
 Q^4-294 Q^2+17\right)
\nonumber\\
&\qquad +\left(Q-\beta _1\right){}^2 \big(-2 \beta _1 \left(2 Q-\beta
 _1\right) \left(1326 Q^2+578 \beta _1 \left(\beta _1-2
 Q\right)-1159\right)
\nonumber\\
&\qquad +1521 Q^4-3318 Q^2+271\big)\bigg)~,
\\
D_3 &= \frac{Q-\beta _1}{34992}\bigg(-324 \beta _0^6 \left(18 \beta _1^2+18 Q^2-36 \beta _1
 Q-11\right)
 +972 \beta _0^5 Q \left(18 \beta _1^2+18 Q^2-36 \beta _1
 Q-11\right)
\nonumber\\
&\qquad +18 \beta _0^4 \big(1836 \beta _1^4+2 \beta _1^2 \left(5157
 Q^2-2536\right)+\beta _1 \left(10144 Q-5940 Q^3\right)
\nonumber\\
&\qquad \qquad +1134 Q^4-5073
 Q^2-7344 \beta _1^3 Q+2135\big)
\nonumber\\
&\qquad-36 \beta _0^3 Q \big(1836 \beta _1^4-8 \beta _1 \left(945
 Q^2-1268\right) Q
 +4 \beta _1^2 \left(2781 Q^2-1268\right)
\nonumber\\
&\qquad \qquad +1944 Q^4-5568 Q^2-7344 \beta
 _1^3 Q+2135\big)
\nonumber\\
&\qquad -18 \beta _0^2 \big(3468 \beta _1^6-2 \beta _1 \left(8577
 Q^4-31690 Q^2+19469\right) Q+6 \beta _1^4 \left(8534
 Q^2-2621\right)
\nonumber\\
&\qquad\qquad  +\beta _1^3 \left(62904 Q-66096 Q^3\right)+\beta _1^2
 \left(46929 Q^4-94594 Q^2+19469\right)
\nonumber\\
&\qquad \qquad +2457 Q^6-15519 Q^4+23929
 Q^2-20808 \beta _1^5 Q-3563\big)
\nonumber\\
&\qquad +18 \beta _0 Q \big(3468 \beta _1^6+6 \beta _1^4 \left(8840
 Q^2-2621\right)+\beta _1^3 \left(62904 Q-73440 Q^3\right)
\nonumber\\
&\qquad \qquad +\beta _1^2
 \left(58215 Q^4-99666 Q^2+19469\right)+\beta _1 \left(-25038 Q^5+73524
 Q^3-38938 Q\right)
\nonumber\\
&\qquad \qquad +4563 Q^6-21186 Q^4+26064 Q^2-20808 \beta _1^5
 Q-3563\big)
\nonumber\\
&\qquad +39304 \beta _1^8-314432 \beta _1^7
 Q+68 \beta _1^6 \left(16439 Q^2-3426\right)-136 \beta _1^5 \left(16949
 Q^2-10278\right) Q
\nonumber\\
&\qquad +\beta _1^4
 \left(3013930 Q^4-3630270 Q^2+419196\right)
\nonumber\\
&\qquad -8 \beta
 _1^3 Q\left(319753 Q^4-650295 Q^2+209598\right)
\nonumber\\
&\qquad +\beta _1^2 \left(1376037 Q^6-4323945 Q^4+2863311
 Q^2-112447\right)
\nonumber\\
&\qquad +\beta _1 Q\left(-428922 Q^6+1970658 Q^4-2373054 Q^2+224894
 \right)
\nonumber\\
&\qquad +59319 Q^8-383643 Q^6+787221 Q^4-133687 Q^2+1890\bigg)~.
\label{eq:D3}
\end{align}

Now, we compare the above partition function of $(A_1,D_3)$ 
with that of $(A_1,A_3)$
\begin{align}
 \mathcal{Z}_{(A_1,A_3)} = \langle 0 |I^{(3)}\rangle~,
\label{eq:ZA1A3}
\end{align}
the latter of which was evaluated in \cite{Nishinaka:2019nuy}. While they are realized by two different
class $\mathcal{S}$ constructions, the $(A_1,A_3)$ and
$(A_1,D_3)$ theories are the identical AD theory \cite{Cecotti:2010fi, Xie:2012hs}. Therefore, \eqref{eq:ZA1D3}
and \eqref{eq:ZA1A3} are expected to be identical.
Indeed, one can explicitly check that, under the identification
\begin{align}
 \beta_0 = \frac{\alpha-Q}{2}~,\qquad \beta_1 = \frac{\alpha - 2\beta_2 +3Q}{2}~,
\end{align}
the expansion coefficients
\eqref{eq:D1}--\eqref{eq:D3} that we have evaluated for $\mathcal{Z}_{(A_1,D_3)}$ are precisely
identical to Eqs.~(3.45) -- (3.47) of \cite{Nishinaka:2019nuy}. This means
that, with the further identification
\begin{align}
 -\frac{3\Lambda_3}{4c_1^3} = \frac{3c_3^2}{c_2^3}~,
\end{align}
 \eqref{eq:ZA1D3} is indeed identical to \eqref{eq:ZA1A3} up to a
prefactor and higher orders of the expansion that we have not evaluated here. This is a strong evidence for our
ansatz \eqref{ansatz3/2} for $|I^{(3/2)}\rangle$.

Finally, we comment on the connection of our
$(A_1,D_3)$ partition function to the Painlev\'e II. It is
shown in
\cite{Nishinaka:2019nuy} that the partition function of $(A_1,A_3)$
evaluated as $\langle 0|I^{(3)}\rangle$ is identical to the series
expression for the tau function
of the Painlev\'e II evaluated in \cite{Bonelli:2016qwg}. The
identification of \eqref{eq:ZA1D3} and \eqref{eq:ZA1A3} then implies
that the same tau function is also reproduced as $\langle \Delta
|I^{(3/2)}\rangle$. Indeed, the irregular conformal block $\langle
\Delta |I^{(3/2)}\rangle$ was essentially evaluated via a different
method in \cite{Nagoya:2018} and shown to be identified as the power
series expression for the tau function evaluated in
\cite{Bonelli:2016qwg}. As we will comment in the next section, it would be
interesting to study how the construction of the irregular conformal
block in \cite{Nagoya:2018} and the one we have discussed here are related.

\section{Summary and discussions}
\label{sec:conclusion}

In this paper, we have conjectured a set of differential equations that
the Liouville irregular state $|I^{(r)}\rangle$ of a general
half-integer rank $r$ satisfies.
This extends the generalized AGT correspondence to all the 
$(A_1,A_\text{even})$ and $(A_1,D_\text{odd})$ types Argyres-Douglas
theories. As shown in Sec.~\ref{sec:lower-ranks}, for lower
half-integer ranks, our conjecture has been verified by deriving it as a 
suitable
limit of a similar set of differential equations for integer
ranks. This limit is interpreted as the 2D
counterpart of a 4D RG-flow from $(A_1,D_{2n})$ to $(A_1,D_{2n-1})$. For
rank $5/2$, our conjectured equations are equivalent to those recently discovered
in the pioneering work \cite{Poghosyan:2023zvy}.

For
rank $3/2$, we have solved our conjectured equations and find a
power series expression for the irregular state
$|I^{(3/2)}\rangle$. Here, we have expanded $|I^{(3/2)}\rangle$ in terms of generalized
descendants of the rank-one state $|I^{(1)}\rangle$, which can be
regarded as a
straightforward generalization of the ansatz used in
\cite{Poghosyan:2023zvy} for $|I^{(5/2)}\rangle$. Inspired by these
results, we conjecture that the rank-$(n-\frac{1}{2})$ irregular state
$|I^{(n-\frac{1}{2})}\rangle$ is always expanded as in
\eqref{eq:ansatz_n-1/2} for any positive integer $n$.\footnote{Note that this is also consistent with
the expansion of the rank-$1/2$ state studied in \cite{Gaiotto:2009ma}.}

There are clearly many future directions, some of which we list below:
\begin{itemize}
 \item It would be nice to prove that our conjectured equations
       \eqref{eq:diff-eq-halfint} are consistent with $[L_k,L_m] =
       (k-m)L_{k+m}$ for $k,m\geq 0$ for all positive integer $n$, and
       therefore give a representation of the Virasoro algebra for all
       half-integer ranks. We have only checked this for
       $n=1,2,3,\cdots,11$.
 \item It would be desirable to solve the recursive equation
       \eqref{ourrep} to obtain a closed form expression for
       $f_k(\Lambda_{n},\cdots,\Lambda_{2n-1})$ for a general positive
       integer $n$. 
 \item One can solve our conjectured equations
       \eqref{eq:diff-eq-halfint} for all positive integers $n$, which
       is expected to lead to a power series expression for
       $|I^{(r)}\rangle$ for an arbitrary half-integer rank $r$. Using
       these expressions, one can then compute the partition functions
       of all $(A_1,A_\text{even})$ and $(A_1,D_\text{odd})$ theories.
 \item In \cite{Kimura:2022yua}, the authors evaluated the Nekrasov partition function of the
       $(A_2,A_5)$ theory, i.e.,
       $SU(2)$ gauge theory coupled to $(A_1,D_3)$, $(A_1,D_6)$
       and a fundamental hypermultiplet. Since differential
       equations for the rank-$3/2$ irregular
       state were not identified at that time, the results of
       \cite{Kimura:2022yua} are limited to the case in which the relevant
       coupling and the VEV of Coulomb branch operators of the
       $(A_1,D_3)$ sector are turned off. Given the differential
       equations \eqref{eq:diff-eq-halfint}, one can now extend the
       results of \cite{Kimura:2022yua} to more general cases.
 \item Up to a conformal transformation, $\langle \Delta
       |I^{(n-\frac{1}{2})}\rangle$ is equivalent to the irregular
       conformal block $\langle 0| \Phi_\Delta(z) |I^{(n-\frac{1}{2})}\rangle$,
       where $\Phi_\Delta (z)$ is a vertex operator corresponding to a regular singularity. In
       \cite{Nagoya:2018}, a similar vertex operator is constructed as a
       linear map from $M_{\Lambda}^{[n]}$ to $M_{\Lambda'}^{[n]}$,
       where $M_{\Lambda}^{[n]}$ is a
       Virasoro-module spanned by
       $L_{-i_1+n}L_{-i_2+n}\cdots
       L_{-i_k+n}|\Lambda\rangle$ ($i_1\geq \cdots \geq
       i_k>0$) with $|\Lambda\rangle$ satisfying $L_k|\Lambda\rangle =
       \Lambda_k|\Lambda\rangle$ for $n\leq k\leq 2n-1$ and $L_{k\geq
       2n} |\Lambda\rangle = 0$. While our $|I^{(n-\frac{1}{2})}\rangle$
       is defined in the usual Verma module of the Virasoro algebra with
       a regular highest-weight state, it
       would be interesting to study its relation to the results of \cite{Nagoya:2018}.
\end{itemize}



\section*{Acknowledgements}

We are grateful to Yasuyuki Hatsuda, Katsushi Ito, Kazunobu Maruyoshi, and Sanefumi Moriyama for helpful
discussions. T.~Nishinaka also thanks Takuya Kimura for helpful
discussions in a separate but related collaboration.
The authors' research is partially supported by JSPS
KAKENHI Grant Number JP21H04993.
In addition, T.~Nakanishi's research is partially supported by JST
Program ``The
Establishment of University Fellowships Towards the Creation of Science
Technology Innovation'' Grant Number JPMJFS2138,
and T.~Nishinaka's research is partially supported by
JSPS KAKENHI Grant Numbers JP18K13547, 23K03394 and 23K03393.

\begin{appendices}
 
\section{Expressions for $f_i(\Lambda_n,\Lambda_{n+1},\cdots,\Lambda_{2n-1})$}
\label{app:f}

Our conjectured equations  \eqref{eq:diff-eq-halfint} contain functions $f_i$ of
$\Lambda_n,\Lambda_{n+1},\cdots,\Lambda_{2n-2}$ and
$\Lambda_{2n-1}$. As we explain in the main text, these functions are recursively
defined by \eqref{ourrep}.
Here, we list explicit expressions for
$f_i(\Lambda_n,\Lambda_{n+1},\cdots,\Lambda_{2n-1})$ for lower values of
$n$. 

\subsection{$n=1$ (rank $1/2$)}

\begin{align}
 f_0(\Lambda_1) = 0~.
\end{align}

\subsection{$n=2$ (rank $3/2$)}

\begin{align}
f_1(\Lambda_2,\Lambda_3) &= \frac{(\Lambda_2)^2}{2\Lambda_3}~,\qquad  f_0(\Lambda_2,\Lambda_3) = 0~.
\end{align}

\subsection{$n=3$ (rank $5/2$)}

\begin{align}
 f_2(\Lambda_3,\Lambda_4,\Lambda_5) &= -\frac{(\Lambda_4)^3}{3(\Lambda_5)^2}
 + \frac{\Lambda_3\Lambda_4}{\Lambda_5}~,
\\
 f_1(\Lambda_3,\Lambda_4,\Lambda_5) &= \frac{(\Lambda_4)^4}{3(\Lambda_5)^3} -
 \frac{\Lambda_3(\Lambda_4)^2}{(\Lambda_5)^2} +
 \frac{(\Lambda_3)^2}{\Lambda_5}~,
\\[1mm]
 f_0(\Lambda_3,\Lambda_4,\Lambda_5) &= 0~.
\end{align}

\subsection{$n=4$ (rank $7/2$)}

\begin{align}
 f_3(\Lambda_4,\Lambda_5,\Lambda_6,\Lambda_7) &=
 \frac{(\Lambda_6)^4}{4(\Lambda_7)^3} -
 \frac{\Lambda_5(\Lambda_6)^2}{(\Lambda_7)^2} +
 \frac{(\Lambda_5)^2 + 2\Lambda_4\Lambda_6}{2\Lambda_7}~,
\\
f_2(\Lambda_4,\Lambda_5,\Lambda_6,\Lambda_7) &=
 -\frac{(\Lambda_6)^5}{2(\Lambda_7)^4} +
 \frac{2\Lambda_5(\Lambda_6)^3}{(\Lambda_7)^3} -
 \frac{2(\Lambda_5)^2\Lambda_6}{(\Lambda_7)^2} -
 \frac{\Lambda_4(\Lambda_6)^2}{(\Lambda_7)^2} +
 \frac{2\Lambda_4\Lambda_5}{\Lambda_7}~,
\\
f_1(\Lambda_4,\Lambda_5,\Lambda_6,\Lambda_7) &=
 \frac{(\Lambda_6)^6}{2(\Lambda_7)^5} -
 \frac{9\Lambda_5(\Lambda_6)^4}{4(\Lambda_7)^4} +
 \frac{(\Lambda_6)^2\Big(3(\Lambda_5)^2 + \Lambda_4\Lambda_6
 \Big)}{(\Lambda_7)^3} 
\nonumber\\
&\qquad - \frac{\Lambda_5\Big((\Lambda_5)^2 +
 6\Lambda_4\Lambda_6\Big)}{2(\Lambda_7)^2} +
 \frac{3(\Lambda_4)^2}{2\Lambda_7}~,
\\
f_0(\Lambda_4,\Lambda_5,\Lambda_6,\Lambda_7) &=0~.
\end{align}

\subsection{$n=5$ (rank $9/2$)}

\begin{align}
f_4(\Lambda_5,\cdots,\Lambda_9) &= -\frac{(\Lambda_8)^5}{5(\Lambda_9)^4} +
 \frac{\Lambda_7(\Lambda_8)^3}{(\Lambda_9)^3} -
 \frac{\Lambda_8\left((\Lambda_7)^2 +
 \Lambda_6\Lambda_8\right)}{(\Lambda_9)^2} + \frac{\Lambda_6\Lambda_7 + \Lambda_5\Lambda_8}{\Lambda_9}~,
\\[4mm]
f_3(\Lambda_5,\cdots,\Lambda_9) &= \frac{3(\Lambda_8)^6}{5(\Lambda_9)^5} -
 \frac{3\Lambda_7(\Lambda_8)^4}{(\Lambda_9)^4} +
 \frac{2(\Lambda_8)^2\left(2(\Lambda_7)^2 +
 \Lambda_6\Lambda_8\right)}{(\Lambda_9)^3} 
\nonumber\\
&\qquad - \frac{(\Lambda_7)^3 + 4\Lambda_6\Lambda_7\Lambda_8 +
 \Lambda_5(\Lambda_8)^2}{(\Lambda_9)^2} + \frac{(\Lambda_6)^2 + 2\Lambda_5\Lambda_7}{\Lambda_9}~,
\\[4mm]
f_2(\Lambda_5,\cdots, \Lambda_9)&= -\frac{(\Lambda_8)^7}{(\Lambda_9)^6} +
 \frac{27\Lambda_7(\Lambda_8)^5}{5(\Lambda_9)^5} -
 \frac{3(\Lambda_8)^3\left(3(\Lambda_7)^2 +
 \Lambda_6\Lambda_8\right)}{(\Lambda_9)^4} 
\nonumber\\
&\qquad + \frac{\Lambda_8\left(4(\Lambda_7)^3 + 9\Lambda_6\Lambda_7\Lambda_8 +
 \Lambda_5(\Lambda_8)^2\right)}{(\Lambda_9)^3}
\nonumber\\
&\qquad - \frac{3\left(\Lambda_6(\Lambda_7)^2 + (\Lambda_6)^2\Lambda_8 +
 \Lambda_5\Lambda_7\Lambda_8\right)}{(\Lambda_9)^2} + \frac{3\Lambda_5\Lambda_6}{\Lambda_9}~,
\\[4mm]
f_1(\Lambda_5,\cdots,\Lambda_9) &= \frac{(\Lambda_8)^8}{(\Lambda_9)^7} -
 \frac{6\Lambda_7(\Lambda_8)^6}{(\Lambda_9)^6} +
 \frac{4(\Lambda_8)^4\left(15(\Lambda_7)^2+4\Lambda_6\Lambda_8\right)}{5(\Lambda_9)^5}
\nonumber\\
&\qquad  -\frac{(\Lambda_8)^2\left(8(\Lambda_7)^3
 + 12\Lambda_6\Lambda_7\Lambda_8 +
 \Lambda_5(\Lambda_8)^2\right)}{(\Lambda_9)^4} 
\nonumber\\
&\qquad
 + \frac{(\Lambda_7)^4 +
 8\Lambda_6(\Lambda_7)^2\Lambda_8 + 4(\Lambda_6)^2(\Lambda_8)^2 +
 4\Lambda_5\Lambda_7(\Lambda_8)^2}{(\Lambda_9)^3}
\nonumber\\
&\qquad -
 \frac{2\left((\Lambda_6)^2\Lambda_7 + \Lambda_5(\Lambda_7)^2 +
 2\Lambda_5\Lambda_6\Lambda_8\right)}{(\Lambda_9)^2} +
 \frac{2(\Lambda_5)^2}{\Lambda_9}~,
\\
 f_0(\Lambda_5,\cdots,\Lambda_9) &= 0~.
\end{align}

\end{appendices}

\bibliography{AGT}
\bibliographystyle{utphys}

\end{document}